\documentclass[a4paper,10pt]{article}
\usepackage{jheppub,slashed,color}
\pdfoutput=1
\newcommand{\beq}{\begin{eqnarray}}
\newcommand{\eeq}{\end{eqnarray}}
\newcommand{\be}{\begin{equation}}
\newcommand{\ee}{\end{equation}}
\def\bsp#1\esp{\begin{split}#1\end{split}}
\chardef\MyArticleWithColor=\pdfcolorstackinit page direct{0 g}

\newcommand{\msbar} {\overline{{\rm MS}}}
\def\pythia{{\sc\small Pythia}}
\def\mg{{\sc MG5}\_aMC}
\def\ma{{\sc\small MadAnalysis}}
\def\fj{{\sc\small FastJet}}

\title{LHC constraints and potential on resonant monotop production}

\author[a]{Giacomo~Cacciapaglia}
\author[b]{\!\!, Eric~Conte}
\author[a]{\!\!, Aldo~Deandrea}
\author[c,d]{\!\!, Benjamin~Fuks}
\author[c]{and Hua-Sheng~Shao}

\affiliation[a]{Univ. Lyon, Universit{\' e} Claude Bernard Lyon 1, CNRS/IN2P3, UMR5822 IPNL, F-69622, Villeurbanne, France}
\affiliation[b]{
  Institut Pluridisciplinaire Hubert Curien/D\'epartement Recherches
  Subatomiques, Universit\'e de Strasbourg/CNRS-IN2P3,
  23 Rue du Loess, F-67037 Strasbourg, France}
\affiliation[c]{Laboratoire de Physique Th\'eorique
  et Hautes \'Energies (LPTHE), UMR 7589, Sorbonne Universit\'e et CNRS, 4 place Jussieu, 75252 Paris Cedex 05, France}
\affiliation[d]{Institut Universitaire de France, 103 boulevard Saint-Michel,
  75005 Paris, France}

\emailAdd{g.cacciapaglia@ipnl.in2p3.fr}
\emailAdd{eric.conte@iphc.cnrs,fr}
\emailAdd{deandrea@ipnl.in2p3.fr}
\emailAdd{fuks@lpthe.jussieu.fr}
\emailAdd{huasheng.shao@lpthe.jussieu.fr}

\abstract{
We discuss the phenomenology associated with a resonant monotop collider signal,
{\it i.e.}~a signal in which a
single top quark is resonantly produced in association with missing energy
through an $s$-channel scalar exchange. We study both the bounds originating
from dedicated monotop searches performed by the ATLAS and CMS experiments, and
the constraints associated with other processes that could be induced by a
new physics context favouring monotop production at colliders. The latter class
of constraints includes, in particular, the recasting of analyses from the
LHC and the TeVatron. All theoretical calculations are performed at the
next-to-leading order accuracy in QCD, and we finally combine all results
to establish the present limits on the parameter space and test the relevance of
the monotop signal at the LHC Run 2.
}

\begin{document}
\maketitle
\flushbottom

\section{Introduction}
\label{sec:intro}
Monotop production at colliders consists in the prodution of a single top quark
in association with a large amount of missing transverse energy. This quite
peculiar final state has been investigated at the LHC by both the ATLAS and CMS
collaborations, and both at Run 1 and 2. As monotop production is heavily
suppressed in the Standard Model, its observation would consist in a
clear sign of physics Beyond the Standard Model. In a new physics context, there
exist two main different monotop production mechanisms~\cite{Andrea:2011ws,%
Agram:2013wda, Boucheneb:2014wza}. In the first of them, the monotop system is
produced from a coloured scalar or vector resonance that decays into a top quark
and an invisible neutral fermion, whereas in the second of them, monotops arise
from the production of a single top quark in association with an invisible
scalar or vector boson via the flavour-changing couplings of the latter to the
top and light quarks. After imposing invariance under the full Standard Model
gauge symmetry and invoking simplicity, it can be shown that only scalar
resonant monotop production and vector flavour-changing monotop production are
consistent~\cite{Boucheneb:2014wza}. Whilst several existing
experimental~\cite{Aaltonen:2012ek,Aad:2014wza,Khachatryan:2014uma,%
Sirunyan:2018gka} and phenomenological~\cite{Kamenik:2011nb,Alvarez:2013jqa,%
Agram:2013wda,Boucheneb:2014wza,DHondt:2015nat} studies focus on the
flavour-changing option, the possibility of resonant production has been less
studied, at least comprehensively~\cite{Wang:2011uxa,Agram:2013wda,Aad:2014wza,%
Sirunyan:2018gka}. On different grounds, monotop signatures have also been
considered in the case of compressed supersymmetry~\cite{Fuks:2014lva,%
Hikasa:2015lma,Goncalves:2016tft,Goncalves:2016nil,Duan:2016vpp} and models with
vector-like fermions~\cite{Goncalves:2017soe} or explaining neutrino
masses~\cite{Ng:2014pqa}.

In this work, we reconsider the resonant production of monotop systems via an
intermediate coloured scalar, which consists in the simplest model featuring
monotop production as a key new physics signal and that is allowed after
imposing invariance under the Standard Model gauge symmetry group~\cite{%
Boucheneb:2014wza}. In practice, we embed the generic
effective Lagrangian for resonant monotop production~\cite{Andrea:2011ws} within
the full gauge symmetry requirements of the Standard Model, which severely
constrains the couplings and quantum numbers of the mediator. Hence the coloured
scalar mediator has an electric charge of $2/3$ and consists in a colour
triplet $\sigma$ that couples to a pair of different-flavour right-handed
down-type quarks. Single mediator production, therefore, occurs via these di-quark
couplings, while the decay of the mediator into a (right-handed) top quark plus an
invisible neutral fermion $\chi$ occurs through an independent coupling
parameter. In order for the model to stay monotop-motivated, it is crucial that
the fermion $\chi$ remains undetected when produced, and thus decays outside the
detector. Remarkably, this model resembles a supersymmetry-inspired simplified
model in which the Standard Model field content is supplemented by neutralino and a
right-handed top squark featuring $R$-parity violating couplings to the
down-type quark sector.

Besides the direct investigation of monotop
probes, this model is also constrained by many other searches for new
physics that thus already limit the available parameter space. Hence we will 
take into account searches from LHC Run 1 and 2 involving jets and top quarks in
the final state, as well as dijet searches at the LHC and the TeVatron.
Moreover, constraints on the decay length of the invisible (unstable) particle
produced in association with the top quark and contributions to the top width
also importantly restricts the parameter space. Another important point
concerning the resonant production of a coloured spin-0 boson is that, being a
QCD process, next-to-leading order (NLO) effects are expected to be important.
As the tools allowing for such a calculation at the Monte Carlo level became
available recently~\cite{Degrande:2014sta}, we employ the full
NLO machinery to study the existing bounds, as well as to establish the LHC
potential at Run 2 to test the simplest phenomenologically viable monotop model.

The paper is organised as follows. In Section~\ref{sec:modelbuilding} we briefly
recall the details of the effective Lagrangian describing resonant monotop
production and of the theoretical framework allowing for numerical Monte Carlo
simulations at the NLO level. In Section~\ref{sec:constraints}, we analyse the
existing bounds coming from different sources. Hence, we reinterpret the results
of stop pair searches, include limits from resonance searches using dijet probes
and the constraints originating from direct monotop searches at the LHC Run 1.
We moreover consider the constraints stemming from the modification of the top
quark width and the fact that the neutral fermion $\chi$ has to be long-lived or
decay invisibly. In section~\ref{sec:LHC}, we collect all current constraints on
the parameter space and discuss the LHC Run~2 potential, which will be relevant
for the ongoing experimental analyses, and hence present our conclusions.

\section{Theoretical framework}\label{sec:modelbuilding}
We consider a general class of scenarios describing the resonant production of
monotop systems at hadronic colliders, in which the key model signature consists
in the production of a top quark in association with missing transverse energy
carried by an invisible Majorana fermion $\chi$ of mass $m_\chi$. Imposing
electroweak gauge invariance, the resulting Lagrangian is quite simple~\cite{%
Andrea:2011ws,Boucheneb:2014wza}. The
only phenomenologically viable and gauge-invariant production mechanism involves
two initial-state right-handed down-type quarks that annihilate into a scalar
field $\sigma$ of mass $m_\sigma$ lying in the triplet representation of the QCD
gauge group, and carrying an electric charge $e_\sigma = 2/3$ in unit of the
positron charge. The $\sigma$
particle then decays into a monotop state consisting of a right-handed top quark
and a $\chi$-particle. This mechanism can be described by a
simplified Lagrangian, to be supplemented to the Standard Model one,
\be
   \Delta {\cal L} =
    D_\mu \sigma^\dag D^\mu \sigma - m_\sigma^2 \sigma^\dag \sigma
    + \frac{i}{2}\bar\chi\slashed{\partial}\chi-\frac12m_\chi\bar\chi\chi
    + \Big[
      \lambda\ \sigma \bar d^c P_R d
    + y\ \sigma \bar t P_L \chi
      + {\rm h.c.} \Big]\ ,
\label{eq:mod} \ee
where all indices are left understood for clarity and $P_{L,R}$ denotes the
left-handed and right-handed chirality projectors. This Lagrangian includes
gauge interactions for all new fields and the Yukawa couplings of the $\sigma$
field to a pair of down-type antiquarks (parameterised by a $3\times 3$ matrix 
$\lambda$, antisymmetric in flavour space) and to the $t\chi$ monotop state
(with a strength $y$).

Aiming at precision predictions for monotop production at NLO in QCD, 
the Lagrangian in Eq.~\eqref{eq:mod} must be consistently
renormalised to absorb all ultraviolet divergences appearing in the virtual
one-loop diagrams. Adopting the on-shell renormalisation scheme, the wave-function
and mass renormalisation constants of the five massless quark fields  vanish
($\delta Z_q = \delta m_q = 0$), whereas those of the massive top quark
($\delta Z_t$ and $\delta m_t$) and $\sigma$ field ($\delta Z_\sigma$ and
$\delta m_\sigma^2$) are given, at the first order in the strong coupling
$\alpha_s$, by
\be\bsp
  \delta Z_t = -\frac{\alpha_s}{3 \pi}
    \bigg[\frac{3}{\bar\epsilon} + 4 - 3 \log\frac{m_t^2}{\mu_R^2}\bigg] &
  \qquad\text{and}\qquad
  \delta m_t = -\frac{\alpha_s m_t}{3 \pi}
     \bigg[\frac{3}{\bar\epsilon} + 4 - 3 \log\frac{m_t^2}{\mu_R^2}\bigg] \ ,\\
  \delta Z_\sigma = 0&
  \qquad\text{and}\qquad
  \delta m_\sigma^2 =
    -\frac{\alpha_s m_\sigma^2}{3\pi} \Big[\frac{3}{\bar\epsilon} +  7 - 
    3\log\frac{m_\sigma^2}{\mu_R^2} \Big]\ .
\esp\ee
In our notation, we denote the regularisation/renormalisation scale by $\mu_R$ and the
ultraviolet-divergent parts of the renormalisation constants are given in terms
of $1/\bar\epsilon = 1/\epsilon - \gamma_E + \log4\pi$,  $\gamma_E$ being the
Euler-Mascheroni constant and the number of spacetime dimensions being
$D=4-2\epsilon$. The Majorana fermionic field $\chi$ is a gauge
singlet, so that it does not need to be renormalised at NLO in $\alpha_s$.
In contrast, the wave-function renormalisation constant of the gluon field
($\delta Z_g$) reads
\be
  \delta Z_g =
    - \frac{\alpha_s}{6 \pi} \bigg[\frac{1}{\bar\epsilon} - \log\frac{m_t^2}{\mu_R^2}\bigg]
    - \frac{\alpha_s}{24 \pi} \bigg[\frac{1}{\bar\epsilon} - \log\frac{m_\sigma^2}{\mu_R^2}\bigg]\ .
\ee
We moreover enforce that the running of the strong coupling constant solely originates
from gluons and \mbox{$N_f=5$} flavours of light quarks and renormalise it by
subtracting, at zero momentum transfer, all contributions from top and $\sigma$
loops. Any effect induced by these massive fields is hence decoupled and
absorbed in the renormalisation of $\alpha_s$,
\be
  \frac{\delta\alpha_s}{\alpha_s} =
    \frac{\alpha_s}{2\pi\bar\epsilon} \bigg[\frac{N_f}{3} - \frac{11}{2}\bigg] +
    \frac{\alpha_s}{6\pi}\bigg[\frac{1}{\bar\epsilon} - \log\frac{m_t^2}{\mu_R^2}\bigg] +
    \frac{\alpha_s}{24\pi}\bigg[\frac{1}{\bar\epsilon} - \log\frac{m_\sigma^2}{\mu_R^2}\bigg] \ .
\ee
Finally, we choose to renormalise the $\lambda$ and $y$ parameters in the
$\msbar$ scheme,
\be
  \frac{(\delta\lambda)_{ij}}{\lambda_{ij}} = - \frac{\alpha_s}{\pi\bar\epsilon}
  \qquad\text{and}\qquad
  \frac{\delta y}{y} = - \frac{\alpha_s}{2 \pi\bar\epsilon}  \ .
\ee
The renormalisation group running for the new physics couplings $\lambda_{ij}$ and $y$ will be performed with the anomalous dimensions as
$\beta_{\lambda_{ij}}=-\frac{\alpha_s}{\pi}$ and $\beta_{y}=-\frac{\alpha_s}{2\pi}$.

In order to handle $2\to 2$ resonant processes at the NLO accuracy in QCD, we
work in the complex-mass scheme~\cite{Denner:1999gp,Denner:2005fg,Denner:2006ic}
where the renormalisation procedure is handled with the complex masses and complex derived parameters.
Complications may consequently arise with the choice of proper Riemann sheets
when the derivation of the various renormalisation constants is at stake~\cite{Frederix:2018nkq}.
However, in our simplified model, there is no $\mathcal{O}(\alpha_s)$
contribution to the particle decay widths at tree level, so that such
complications are avoided. To achieve the NLO QCD accuracy in the whole phase
space, we nevertheless need to evaluate the unstable particle widths by
including NLO QCD corrections. For simplicity, we fix in these calculations
the renormalisation scale $\mu_R$ to the respective particle masses, and include
the renormalisation group running of the $\alpha_s$, $\lambda_{ij}$ and $y$
couplings. In the context of the width calculations, the potential impact of
using a different scale is a pure next-to-NLO effect, and has
therefore been ignored.

\section{Resonant monotops in the LHC era}
\label{sec:constraints}
The model that has been described in Section~\ref{sec:modelbuilding} has six
free parameters, \textit{i.e.} two masses and four couplings,
\be
  \Big\{m_\chi,\ m_\sigma,\ y,\ \lambda_{12} \!=\! -\lambda_{21},\
    \lambda_{23}\!=\! -\lambda_{32},\ \lambda_{31} \!=\! -\lambda_{13}
  \Big\}\ .
\ee
To exhaustively explore its phenomenology, in particular at the LHC, we first
simplify the parameter space by assuming that only the $\lambda_{12}$ parameter
dominates,
\be
  \lambda\equiv\lambda_{12} \gg \lambda_{23}, \lambda_{31} \approx 0 \ .
\ee
Such a choice allows us not only to avoid the flavour constraints that arise in
particular from kaon mixing~\cite{Wang:2011uxa}, but also to maximise the
monotop production cross section at the LHC by virtue of parton density effects.
The monotop signal then solely depends on the scalar mass $m_\sigma$ and the
relative magnitude of the $\lambda$ and $y$ parameters that control the two
branching ratios
\be
 \mbox{BR} (\sigma \to t \chi) \equiv \mbox{BR}_{t\chi} \qquad\text{and}\qquad
 \mbox{BR} (\sigma \to \bar d \bar s) \equiv \mbox{BR}_{jj} =
    1-\mbox{BR}_{t\chi} \ .
\ee
Our exploration strategy, therefore, consists in slicing the
parameter space for fixed values of the $\mbox{BR}_{t\chi}$ branching ratio and of the coupling $\lambda$, 
and in studying how the constraints evolve for increasing importance of the dijet
decay channel. We therefore describe our model in terms of the four parameters
\be
  \Big\{m_\chi,\ m_\sigma,\ \lambda,\ {\rm BR}_{t\chi} \Big\}\ ,
\ee
where $y$ has been traded with the $\mbox{BR}_{t\chi}$ branching ratio. We have
kept the $\lambda$ parameter free (and not the $y$ one) as it directly controls
the resonant production rate of a $\sigma$ particle. As the aim of this work is
to focus on monotop models, we will exclude from our investigations any
parameter space region in which the $\sigma$ particle cannot decay into a
monotop system, {\it i.e.} regions for which $m_\sigma < m_\chi + m_t$.

In this section, we will consider two classes of constraints, namely those that
are respectively independent and dependent on $\lambda$. The former ones allow us to
directly exclude regions in the $(m_\chi, m_\sigma)$ parameter space at fixed
$\mbox{BR}_{t \chi}$, whilst the latter ones allow us to derive upper limits on
the coupling $\lambda$ for each point of the same mass plane. The first category of
constraints includes typical LHC searches for the production of a pair of
coloured resonances (see Section~\ref{sec:stops}), as well as searches capable
of targeting the doubly-resonant production of a pair of dijet systems (see
Section~\ref{sec:dijetpairs}). These searches are indeed sensitive to the
production and decay of a pair of $\sigma$ particles. On different lines, the
second category of constraints includes bounds that could originate from dijet
(see Section~\ref{sec:dijet}) and monotop (see Section~\ref{sec:monotops})
probes, as these two final states can be induced by the resonant production and
decay of a single $\sigma$ particle. In addition, for parameter space regions in
which $m_\chi < m_t$, the top quark can undergo a three-body decay in a
$\chi j j$ final state via an off-shell $\sigma$ particle. While no direct
search is currently dedicated to such a decay, measurements of the top width
 yield indirect constraints on the model (see Section~\ref{sec:widths}).
The fermion $\chi$ is in principle allowed to decay into a $t^{(*)} j j$ system,
with the final-state top quark being potentially off-shell, and one must ensure
that $\chi$ is stable on LHC detector scales (see also in
Section~\ref{sec:widths}). Finally, in the same section, we  comment on the
usage of the narrow-width approximation for the $\sigma$ particle.

In our phenomenological investigations, we rely on a numerical evaluation in
four dimensions of all loop integrals, so that the numerical results must be
complemented by rational terms related to the $\epsilon$-dimensional components
of the integrals~\cite{delAguila:2004nf,Pittau:2004bc,Xiao:2006vr,Su:2006vs}.
They consist of the
so-called $R_1$ and $R_2$ terms, the former being universal and connected to the
denominators of loop integrands and the
latter being model-dependent and associated with the numerators of loop integrands.
In practice, we perform loop-integral computations with the {\sc MadLoop}
package~\cite{Hirschi:2011pa} that automatically estimates the $R_1$
contributions and makes use of a finite set of special Feynman rules derived
from the bare Lagrangian~\cite{Ossola:2008xq} to estimate the $R_2$ ones. We
have computed those $R_2$ Feynman rules by implementing the Lagrangian of
Eq.~\eqref{eq:mod} into the {\sc FeynRules} package~\cite{Alloul:2013bka}, that
we have jointly used with the NLOCT~\cite{Degrande:2014vpa} and
{\sc FeynArts}~\cite{Hahn:2000kx} programs to export the model under the form of
a UFO module~\cite{Degrande:2011ua}. Such a module contains here, in addition to
tree-level information, all ultraviolet counterterms and $R_2$ Feynman rules
needed for numerical loop-calculations in QCD in the context of our monotop
model. In practice, this has allowed us to
make use of the {\sc MadGraph5}\_aMC@NLO~\cite{Alwall:2014hca} (\mg)
platform for the generation of the  LHC signals at the NLO accuracy in QCD.

Moreover, for the resonant processes in which the complex-mass scheme must be
employed, we have verified that the NLO widths computed by \mg\  agree
with independent in-house calculations. We now present the current bounds
on the parameter space of the model.

\subsection{LHC constraints on $\sigma$ pair-production from supersymmetry
searches}
\label{sec:stops}

The $\sigma$ particle, being charged under the QCD gauge group, can be
copiously pair-produced at the LHC. Considering a decay mode in which both
$\sigma$ particles decay into a monotop system,
\be
  p p \to \sigma\sigma^\dag \to t\chi\  \bar t \chi\ ,
\label{eq:stop1}\ee
one obtains a signature that could be probed by typical stop searches in the
top-antitop plus missing transverse energy ($t\bar t + \slashed{E}_T$) channel
as well as by generic supersymmetry searches in the jets plus missing energy
mode (assuming that both final-state top quarks decay hadronically). This latter
class of searches is also sensitive to signals arising from a mixed decay mode in which
one $\sigma$ particle decays into a monotop system and the other one into a
pair of jets:
\be
  p p \to \sigma\sigma^\dag \to t\chi\  j j + {\rm h.c.}
\label{eq:stop2}\ee
The remaining decay option in which both $\sigma$ particles decay into a pair of jets
will be addressed in Section~\ref{sec:dijetpairs}. In the two considered cases
of Eq.~\eqref{eq:stop1} and Eq.~\eqref{eq:stop2}, the
exclusions in the
$(m_\chi, m_\sigma)$ planes only depend on the two masses, with the total rate
being rescaled, respectively, by $\mbox{BR}_{t\chi}^2$ and $2 \mbox{BR}_{t\chi}
(1-\mbox{BR}_{t\chi})$ factors.

\begin{figure}
  \centering
  \includegraphics[width=.6\textwidth]{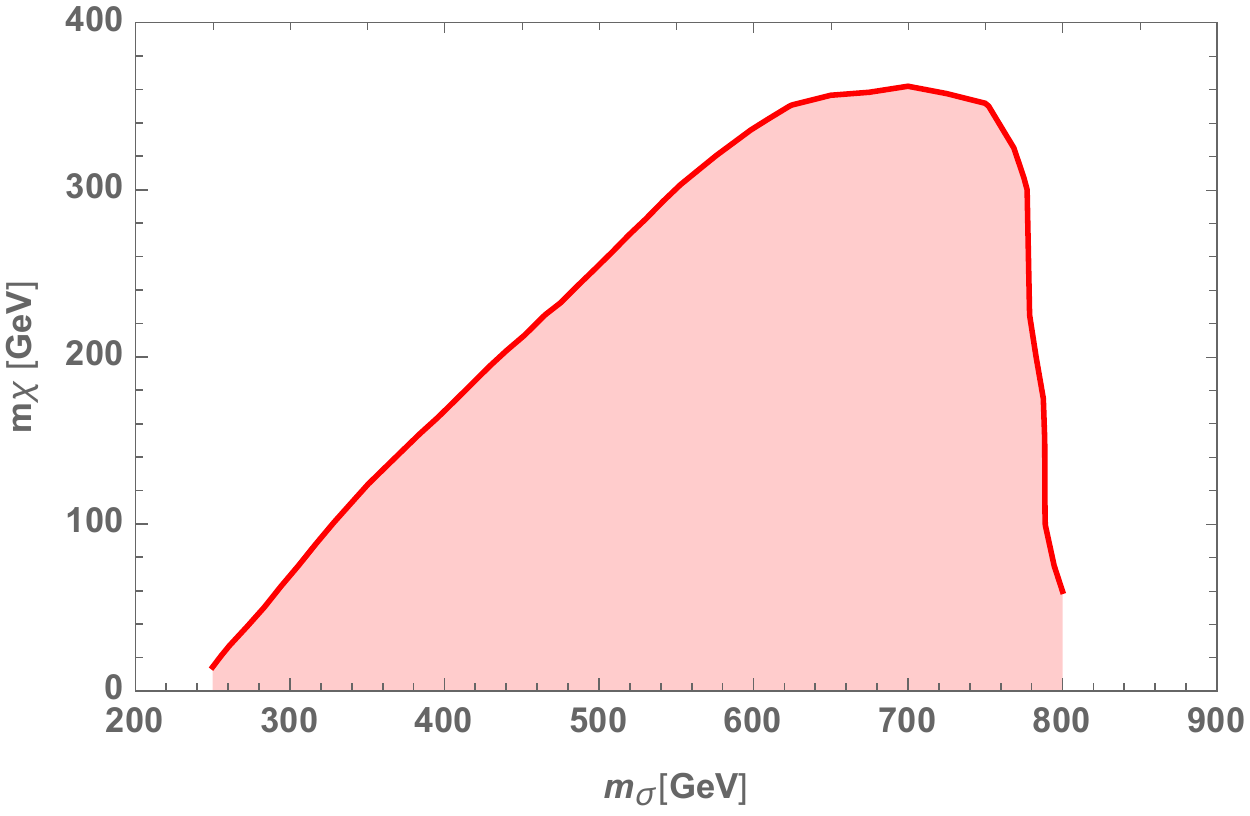}
  \caption{The red region under the curve corresponds to the excluded region at
    95\% confidence level by the reinterpretation of the results of the
    CMS-SUS-17-001 search
    for stops, assuming $\mbox{BR}_{t\chi}=100\%$. Our findings are represented
    in the $(m_\sigma, m_\chi)$ mass plane.}
\label{fig:stoppair}
\end{figure}

As all LHC stop searches give rise to similar bounds, we reinterpret the results
of a single recent search: we thus consider the CMS-SUS-17-001
analysis, which focuses on the dileptonic mode of the top-antitop
system~\cite{Sirunyan:2017leh}. This search is based on 
an integrated luminosity of 35.9~fb$^{-1}$ of LHC Run~2 proton-proton collision
data at a centre-of-mass energy $\sqrt{s}=13$~TeV. It targets a signature
made of two opposite-sign leptons (electrons or muons, with a veto on the
presence of a reconstructed $Z$-boson), light and $b$-tagged jets and a
significant amount of missing transverse momentum. The production of the
hypothetical stop-pair signal, which in our case is identified with the
production of two $\sigma$ scalars decaying as in the process of
Eq.~\eqref{eq:stop1}, is then efficiently separated from the dominant
top-antitop background by a selection on the transverse mass
$m_{T2}$~\cite{Lester:1999tx,Cheng:2008hk} reconstructed from the two
leptons and the missing transverse momentum. In Figure~\ref{fig:stoppair}, we
show the LHC bound for the process in Eq.~\eqref{eq:stop1}, assuming
$\mbox{BR}_{t\chi}=100\%$ and in the plane of the two masses, $(m_\sigma,
m_\chi)$. We made
use of the \ma~5 platform~\cite{Conte:2012fm,Conte:2014zja} and
its public analysis database~\cite{Dumont:2014tja,Conte:2018vmg}, which contains
the corresponding validated reimplementation~\cite{1667773} and
the necessary {\sc Delphes}~3 configuration cards for handling the simulation of
the response of the LHC detectors~\cite{deFavereau:2013fsa}.

We additionally tested the limits arising from two representative ATLAS
searches for dark matter and supersymmetry in the multijet plus missing energy
channel~\cite{Aaboud:2016tnv,Aaboud:2016zdn}. Both these searches target a small
3.2~fb$^{-1}$ luminosity of proton-proton collisions at $\sqrt{s}=13$~TeV and consider a
signature featuring one very hard jet plus subleading hadronic activity. Whilst
the ATLAS-EXOT-2015-03 analysis~\cite{Aaboud:2016tnv} only allows for a
restricted subleading jet activity, the ATLAS-SUSY-2015-06
analysis~\cite{Aaboud:2016zdn} includes signal regions that require a
much more important jet activity. Both these analyses rely on a large set of
signal regions differing by the number of jets, their kinematical configuration
and the amount
of missing transverse energy. Whilst both these ATLAS analyses only consider a
small integrated luminosity of 3.2~fb$^{-1}$ of proton-proton collisions, they are already
limited by the systematics. The resulting bounds are consequently not expected
to improve with an increased LHC luminosity~\cite{Banerjee:2017wxi} and the
subsequent predictions can be robustly used as the best current constraints
originating from multijet plus missing transverse energy production at the LHC.
By using the validated \ma~5 public implementations~\cite{1476800,1510490},
we have found that they only marginally improve the exclusions at the price of
a larger systematic uncertainty due to the fact that the final state in our
model differs from the supersymmetric benchmarks. Thus, we will conservatively
only use the exclusion from the CMS stop search in the following.

\subsection{Searches for a pair of dijet resonances at the LHC}
\label{sec:dijetpairs}

As mentioned in Section~\ref{sec:stops}, the production of a pair of $\sigma$
particles can yield a dijet-pair final state when they both decay  into
quarks,
\be
  p p \to \sigma\sigma^\dag \to jj jj \ .
\ee
Although the relevance of this channel is reduced when monotop production is
large ({\it i.e.}, when $\mbox{BR}_{t\chi}$ is large), it can lead to important
constraints for large $\mbox{BR}_{jj} = (1-\mbox{BR}_{t\chi})$ values as the
resulting rate is proportional to $\mbox{BR}_{jj}^2$.

A new physics signature featuring a pair of dijet systems originating from a
pair of resonances has been searched for at the LHC both during Run~1 and Run~2,
and by both the ATLAS and CMS collaborations. One of the investigated benchmark
models consists of a simplified model where the SM is supplemented
by a light stop decaying into two jets via an $R$-parity violating interaction
with a branching fraction of 100\%. The corresponding experimental results can thus
be directly reinterpreted as a bound on the $\mbox{BR}_{t\chi}$ branching ratio,
as the signal total rate consists of the stop-pair production cross section
rescaled by a $(1-\mbox{BR}_{t\chi})^2$ factor.

We include in our study the CMS-EXO-12-052 final Run~1 search~\cite{Khachatryan:2014lpa}
dedicated to events exhibiting the presence of at least four jets that are then
paired using an algorithm based on their angular
distribution. The discrimination from the leading
multi-jet background is achieved by relying on a set of kinematical variables
including a reduced mass asymmetry between the two pairs of jets. The search
has implemented two signal regions. The first region is dedicated to resonance
masses larger than $300$~GeV and benefits from the full Run~1 dataset with an
integrated luminosity of $19.4$~fb$^{-1}$, whereas the second region is
restricted to a smaller dataset of $12.4$~fb$^{-1}$ and solely considers
resonance masses lying in the [200, 300]~GeV mass window. It has been made
possible to access such low masses thanks to a specific trigger that has been
designed especially for this purpose, with a cost in luminosity.

The same signal can also be constrained with Run~2 data. We reinterpret the
results of a CMS search focusing on low-mass resonances in the [80, 300]~GeV
regime~\cite{CMS:2016pkl} and on a low-luminosity of data ($2.7$~fb$^{-1}$),
and the results of ATLAS analysis of the 2015 and 2016 datasets
($36.7$~fb$^{-1}$)~\cite{Aaboud:2017nmi}. Both Run~2 analyses rely on a boosted
configuration and select events featuring two back-to-back fat jets and a
significant
hadronic activity. After a standard cleaning of the signal by means of various
grooming and pruning methods, the substructure of the fat jet is employed to
improve the quality of the signal and to discriminate it from the background,
together with considerations on the mass asymmetry between the two fat jet
masses and on other kinematical variables.

\begin{figure}
  \centering
  \includegraphics[width=.6\textwidth]{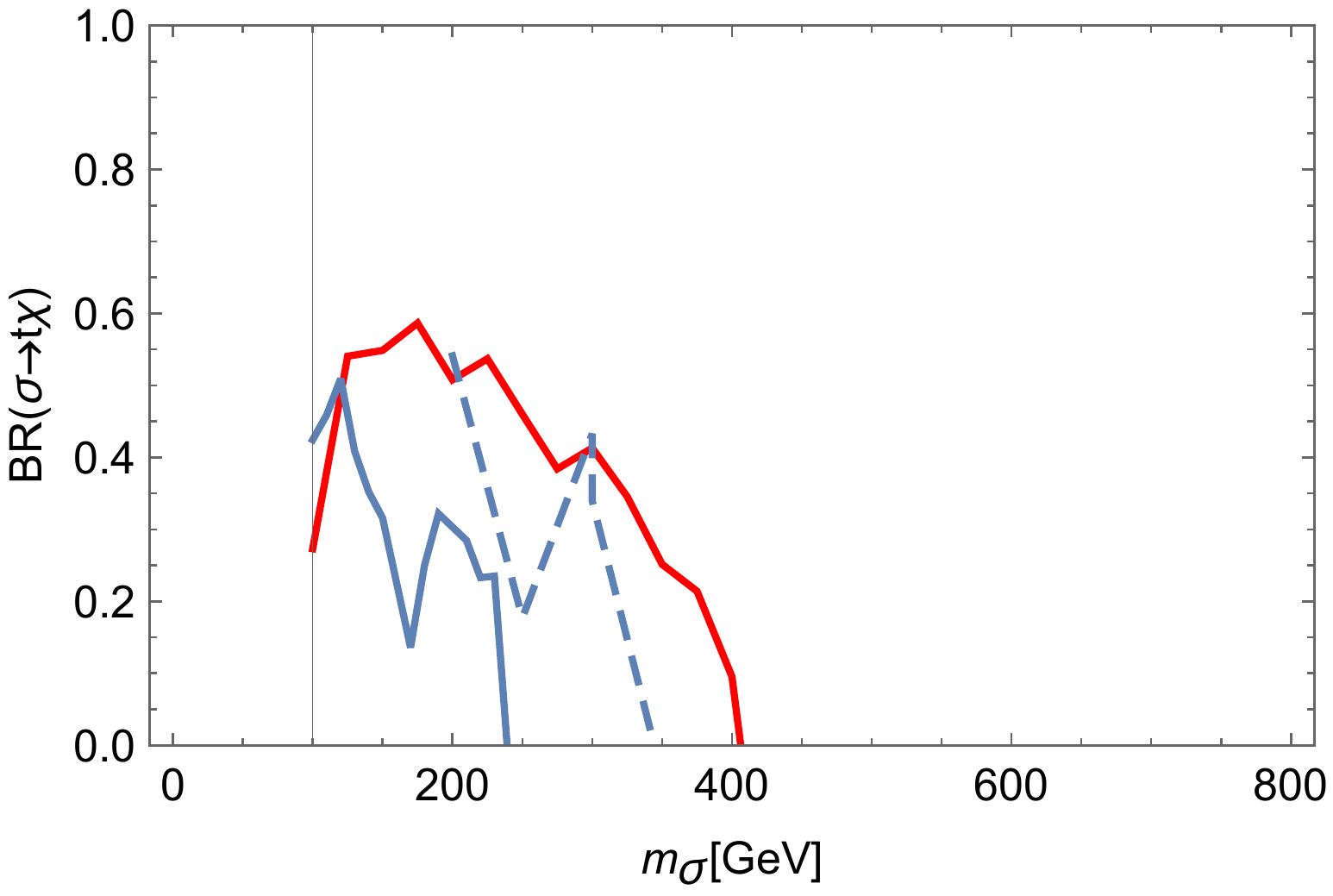}
  \caption{Lower bound on $\mbox{BR}_{t\chi}$ as a function of the mass of
    $m_\sigma$ extracted from the search for a pair of dijet
    resonances in LHC proton-proton data by CMS (blue) at $\sqrt{s}=8$~TeV
    (dashed) and 13~TeV (solid), and by ATLAS (solid red) for $\sqrt{s}=13$~TeV. The area under the curves is the excluded region.
    For the Run~1 CMS search, the discontinuity at $m_\sigma = 300$~GeV stems
    from the transition between two different search strategies respectively
    targeting low-mass and high-mass resonances~\cite{Khachatryan:2014lpa}.}
\label{fig:dijetpair}
\end{figure}

Our results are shown in Figure~\ref{fig:dijetpair}, where we consider the
above-mentioned ATLAS and CMS searches and theoretical estimates of the stop
pair-production cross section evaluated at the NLO accuracy in
QCD~\cite{Degrande:2014sta,Borschensky:2014cia}. The strongest bound arises from
the Run~2 ATLAS search for most values of the $\sigma$ mass, with the exception
of two mass points for which the Run~1 CMS search does a better job and the very
low mass region that benefits from the dedicated CMS Run~2 analysis. The dip
at $m_\sigma = 250$~GeV featured in the CMS Run~1 limit is connected to a small
excess of events in a single bin that has however not been confirmed by ATLAS.
For fixed $\mbox{BR}_{t\chi}$, therefore, a lower bound on the mass $m_\sigma$ can be
extracted.

\subsection{Dijet searches at the TeVatron and the LHC}
\label{sec:dijet}

Dijet searches constrain the single production of the $\sigma$ resonance that
then decays into jets. Due to the large QCD background and because of trigger
requirements, dijet searches at the LHC typically target high invariant masses,
so that dedicated efforts are needed to test the low mass regions that are more
relevant for the monotop model under consideration. At Run 2, the most recent ATLAS
search based on an integrated luminosity of $37$~fb$^{-1}$ has a minimum reach
of $1.1$~TeV~\cite{Aaboud:2017yvp}, while a low-mass dedicated
search~\cite{Aaboud:2018fzt} ($29.3$~fb$^{-1}$) overcomes the trigger limitation
by recording only jet trigger information and reaches invariant masses
as low as $450$~GeV. Similarly, for the CMS analyses of $36$~fb$^{-1}$ of
13~TeV data~\cite{Sirunyan:2018xlo}, the high mass region starts at $1.6$~TeV,
whereas a trigger-based low-mass search allows for reaching down to masses of
600~GeV. A special position is reserved for a very low mass CMS
search~\cite{Sirunyan:2017nvi} with $36$~fb$^{-1}$, where the trigger limitation
is overcome by looking for boosted dijet systems. The boost allows to reduce the
background and, therefore, to reach the $50$--$300$~GeV mass range. Similarly,
the ATLAS collaboration has performed a search with 15.5~fb$^{-1}$ specifically
dedicated to low-mass dijet resonances produced in association with a high-$p_T$
photon~\cite{ATLAS-CONF-2016-070}, which allows for reaching invariant masses in
the $200$--$1200$~GeV range at the price of a lower signal cross section.
At the TeVatron, the most up-to-date search for new particles giving rise to a
dijet final state is from the CDF collaboration~\cite{Aaltonen:2008dn}, where
the signal would arise in our case from the
\be
  p\bar p\rightarrow \sigma/\sigma^{\dagger} \rightarrow jj+X
\ee
process, for $p\bar p$ collisions at $\sqrt{s}=1.96$ TeV. The $95\%$ confidence
level limits exclude resonance mass ranging from 260 GeV to 1400 GeV.

The cross section associated with the production of a dijet system originating
from the decay of a $\sigma$ resonance is proportional to
$\lambda^2(1-{\rm BR}_{t\chi})$, so that for a fixed ${\rm BR}_{t\chi}$ value,
an upper limit on $\lambda$ can be established from the current bounds. In the
numerical recast of the relevant searches, we fixed $m_t = 173.3$~GeV and worked
in the five light quark flavour scheme. In order to estimate the relevant signal
fiducial cross sections and subsequently extract the bounds on the model
parameters, we used the \mg~\cite{Alwall:2014hca} and
\pythia~8.2~\cite{Sjostrand:2014zea} programs to generate NLO-accurate event
samples in which the fixed-order results, obtained by convoluting the NLO
hard-scatering matrix elements with the NLO set of NNPDF~3.0 parton
densities~\cite{Ball:2014uwa}, are matched with parton showers following the
MC@NLO prescription~\cite{Frixione:2002ik}. We then utilised the \ma~5
platform~\cite{Conte:2012fm} and its interface to \fj~\cite{Cacciari:2011ma} to
impose the same jet reconstruction (based on the anti-$k_T$ algorithm~\cite{%
Cacciari:2008gp}) and kinematical cuts as in the experimental
analyses under consideration. In all our computations, the factorisation and
renormalisation scales have been set to the $\sigma$ mass $m_{\sigma}$.

\renewcommand\arraystretch{1.3}%
\begin{table}
  \begin{center}
   \begin{tabular}{|{c}*{3}{|c}|} \hline
     $m_{\sigma}$ [GeV] &
       $\sigma^{\rm NLO,CDF}_{\sigma \to jj}/(\lambda^2(1- BR_{t\chi}))$ [pb] &
       CDF limit [pb]~\cite{Aaltonen:2008dn} \\\hline
     $260$ & $252$ & $110$  \\\hline
     $300$ & $132$ & $45$ \\\hline
     $400$ & $26.1$ & $7.2$ \\\hline
     $500$ & $5.83$ & $3.9$ \\\hline
     $620$ & $0.960$ &  $0.8$ \\\hline
     $700$ & $0.259$ & $0.6$ \\\hline
   \end{tabular}
   \caption{\label{tab:jjXS}Fiducial cross sections for dijet production at the
     TeVatron, in $p\bar{p}$ collisions at $\sqrt{s}=1.96$ TeV, after imposing
     the same jet reconstruction method and signal selection as in the CDF
     analysis of Ref.~\cite{Aaltonen:2008dn}. We compare our (normalised)
     predictions with the CDF limits.}
  \end{center}
\end{table}

\begin{figure}
  \centering
  \includegraphics[width=7cm]{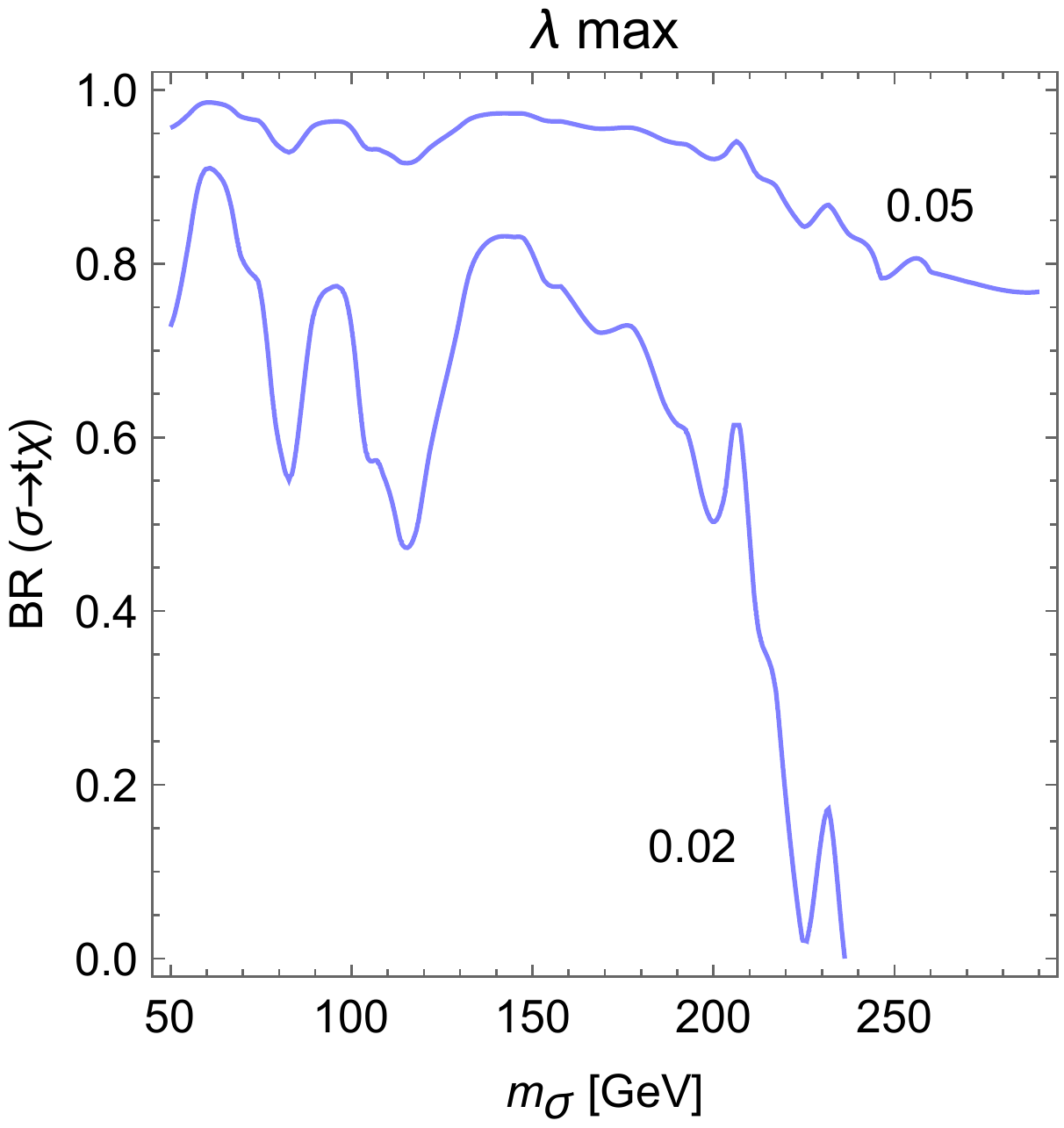}
  \includegraphics[width=7cm]{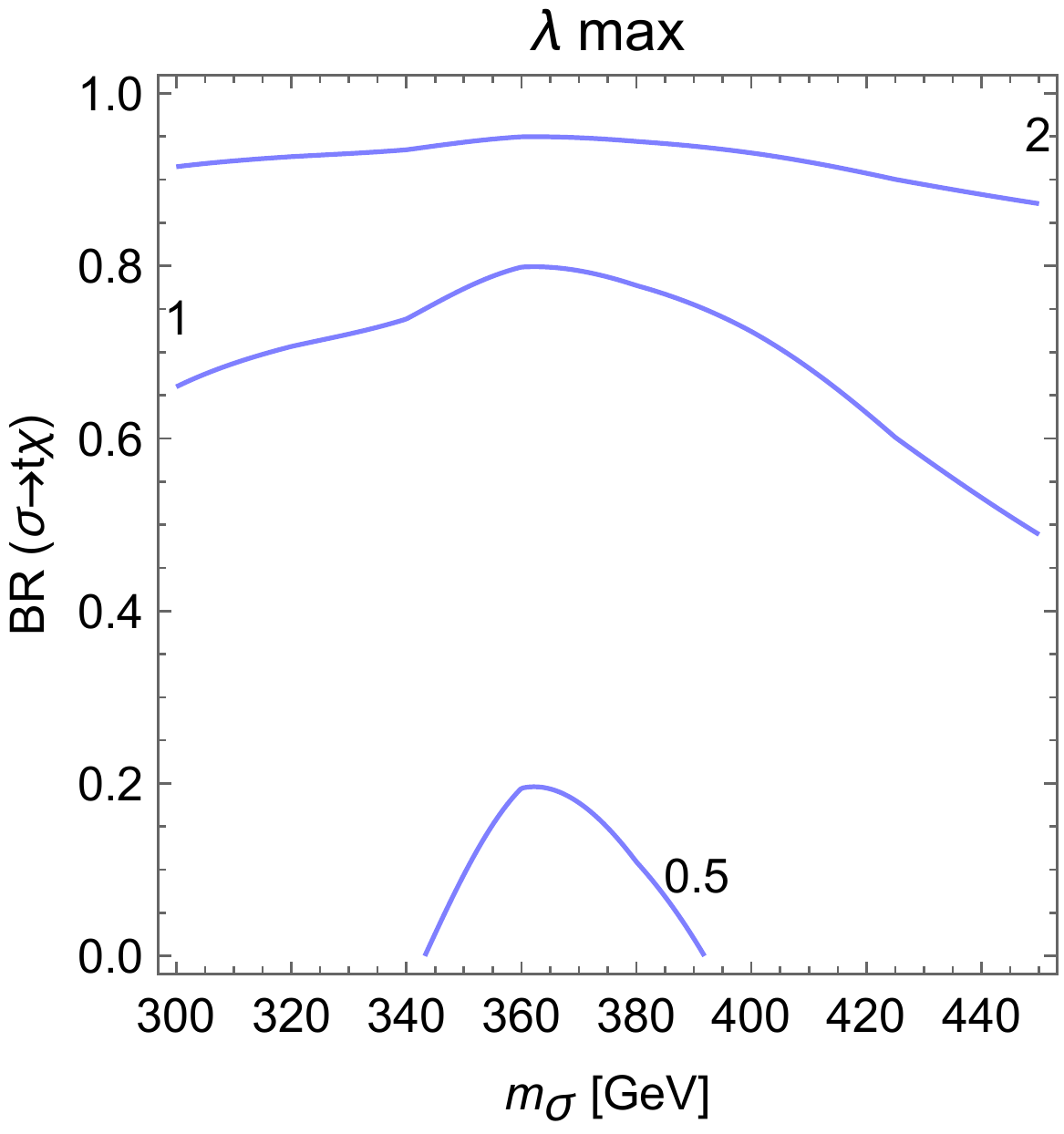}\\
  \includegraphics[width=7cm]{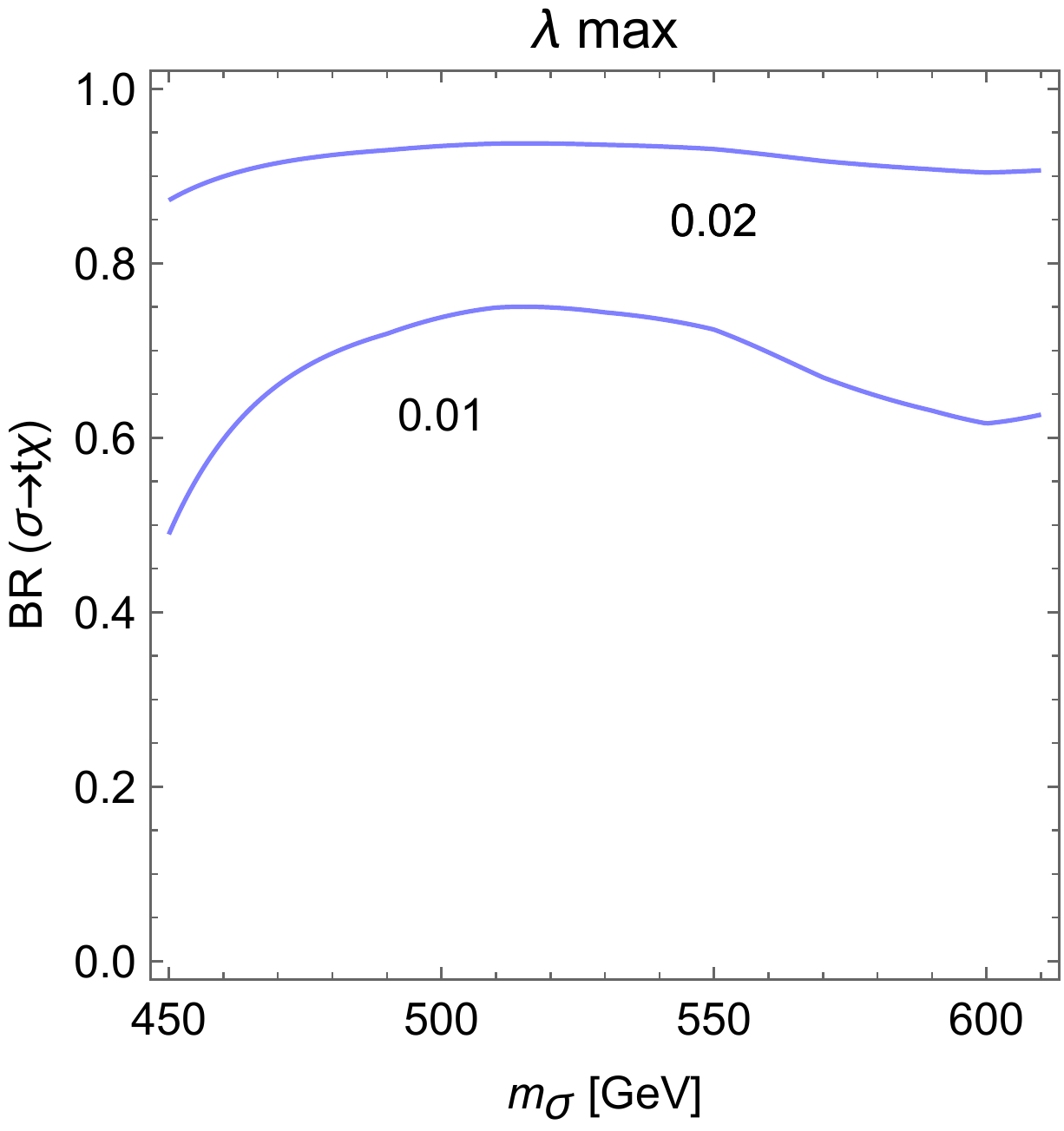}
  \includegraphics[width=7cm]{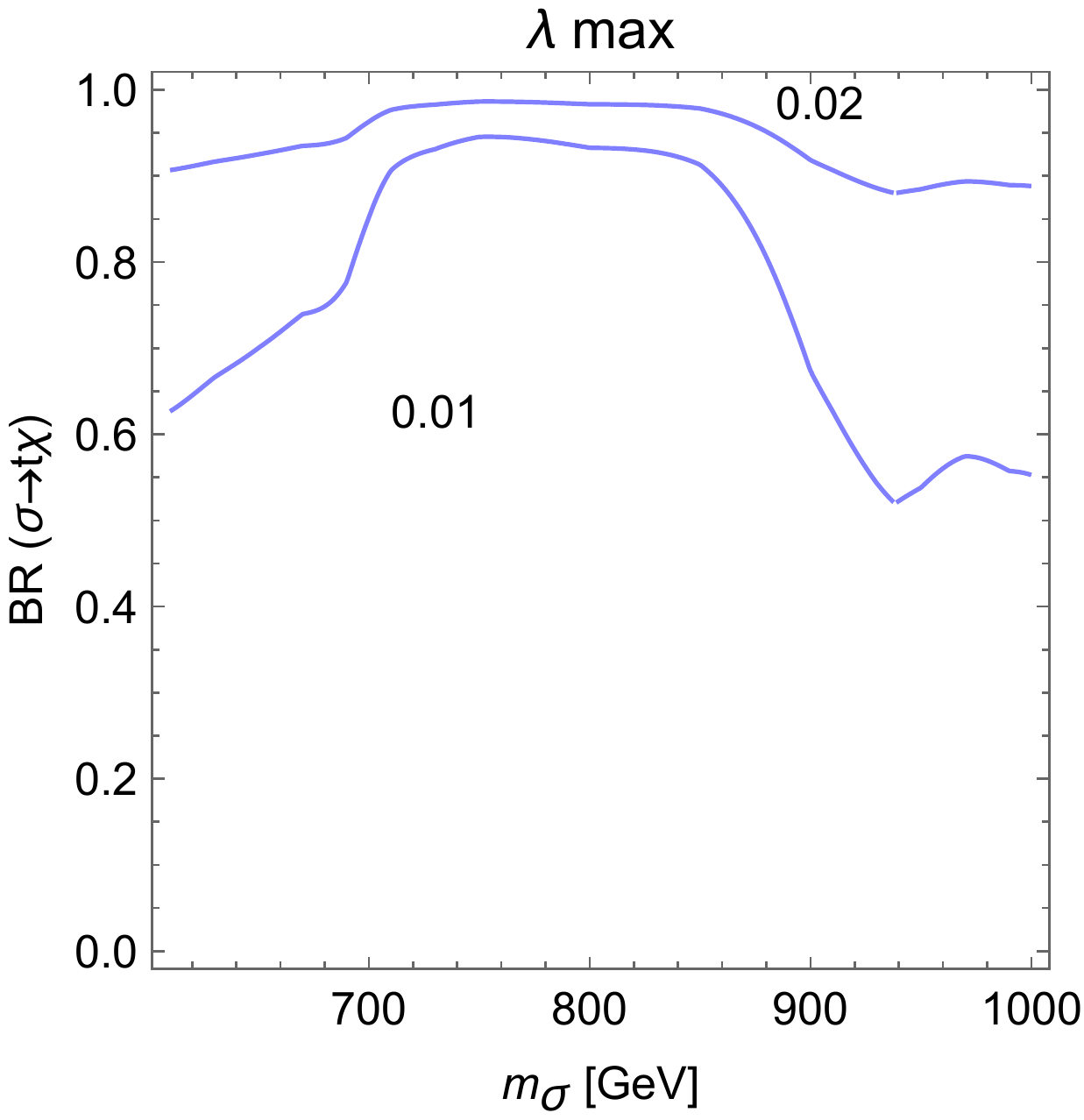}
  \caption{Upper bound on the $\lambda$ parameter denoting the coupling strength
    of the $\sigma$ resonance to down-type quarks as derived from a variety of
    dijet searches at colliders. The results are represented in the $(m_\sigma,
    {\rm BR}_{t\chi})$ plane, and the regions under the curves are excluded for
    $\lambda$ equal to at least the indicated value. In the upper left panel, we
    reinterpreted the low mass CMS search in the boosted regime~\cite{%
    Sirunyan:2017nvi}, whilst on the upper right panel, we considered the CDF
    analysis~\cite{Aaltonen:2008dn}. In the lower left panel, we focused on the
    trigger-based ATLAS analysis~\cite{Aaboud:2018fzt}, and combined it with the
    similar CMS search~\cite{Sirunyan:2018xlo} in the lower right panel,
    recalling that the CMS analysis is insensitive to any $m_\sigma$ value
    smaller than 600~GeV.}
\label{fig:dijet}
\end{figure}

As an example, we focus on the CDF dijet search of Ref.~\cite{Aaltonen:2008dn}. 
In Table~\ref{tab:jjXS} we provide  the predicted NLO fiducial cross sections for
a few selected $m_{\sigma}$ values. Our results can be compared with the CDF
bounds, so that we can extract an upper bound on the $\lambda$ parameter.
Repeating the exercise for all the above-mentioned searches, we present in
Figure~\ref{fig:dijet} the upper limits on $\lambda$ in the $(m_{\sigma},
{\rm BR}_{t\chi})$ plane that we divide in four kinematical configurations in
which different searches dominate.

The low mass regime in which $m_\sigma$ lies in the 50--300~GeV mass window has
been probed by the CMS boosted search of Ref.~\cite{Sirunyan:2017nvi}. The tight
requirements of the trigger are overcome by requiring the presence of at least
one broad jet with $p_T > 500$~GeV. Then, various jet substructure techniques
are employed to discriminate a signal in which the broad jet is issued from a
resonance decaying into a boosted dijet system from the QCD background. The
benchmark model used in the search is a $Z'$ model. Even though our model
contains a scalar resonance and not a vector one, we do not expect significant
kinematical differences in the properties of the dijet system. We have therefore
simply reinterpreted the search in terms of our model by a direct comparison of
the predicted production cross section with the excluded one. The results are
presented in the upper left panel of Figure~\ref{fig:dijet}. The 300--450~GeV
mass window is only covered by the CDF analysis, which implies much weaker
limits on $\lambda$, as shown in the upper right panel of the figure. In the
whole parameter space region, the best upper limit is $\lambda < 0.46$,
{\it i.e.}~one to two orders of magnitude weaker than any limit that could be
derived from the LHC results. For $\sigma$ masses larger than 450~GeV, the
trigger-based low mass search from ATLAS~\cite{Aaboud:2018fzt} kicks in with
limits much stronger than the ones derived from the CDF results, as shown in the
lower left panel of the figure. Finally, for masses greater than 600~GeV, the
ATLAS limits can be combined with those obtained from the trigger-based search
of CMS~\cite{Sirunyan:2018xlo}, which is presented in the lower right panel of
the figure. In summary, we observe that current low mass dijet searches at the
LHC give bounds on the coupling $\lambda$ of order $10^{-2}$, except for the
300--450~GeV mass window where only much weaker CDF limits are applicable.

\subsection{Monotop bounds after Run 1}
\label{sec:monotops}

Monotop searches have been designed to get hints for new physics in a final
state comprised of a single top quark and missing transverse energy. As sketched
in Section~\ref{sec:modelbuilding}, such a final state can originate from the
(potentially resonant) production of a $\sigma$ state followed by its decay into
a $t\chi$ system. The first experimental search for monotops has been carried
out by the CDF collaboration at the TeVatron~\cite{Aaltonen:2012ek} and solely
focused on the flavour-changing monotop production mode. LHC Run~1 data has been
analysed during the last few years, both by the CMS~\cite{Khachatryan:2014uma}
and ATLAS~\cite{Aad:2014wza} collaborations. Whilst the CMS analysis again
focused only on flavour-changing monotop production, the ATLAS results have been
interpreted both in the flavour-changing and resonant scenarios. The way in
which they are presented however makes their reinterpretation in different
theoretical frameworks highly non-trivial. The ATLAS analysis indeed assumes that
all model coupling parameters are equal to a common value, and a bound on this
value is presented in terms of the resonance mass. It is consequently impossible
to make use of the results for model configurations not satisfying this
requirement. The first monotop analysis of the LHC Run~2 results has also been
recently released by the CMS collaboration~\cite{Sirunyan:2018gka}, but it
targets boosted monotop systems so that it is not relevant for the mass scales
probed in this work. For these reasons, we concentrate on the CMS Run~1 monotop
analysis that we consider as representative for the most constraining direct LHC
search on the resonant monotop model.

To this aim, we have implemented the CMS-B2G-12-022 search for monotop
production in proton-proton collisions at a centre-of-mass energy of
8~TeV~\cite{Khachatryan:2014uma} in the \ma~5
framework~\cite{Conte:2012fm,Conte:2014zja,Dumont:2014tja}. We have validated
our reimplementation by making predictions for two monotop scenarios for which
the CMS collaboration has provided signal events and associated official
results. The first of these scenarios concerns flavour-changing monotop
production (with an invisibly-decaying vector boson of 500~GeV and all model
couplings fixed to 0.1), whilst the second one addresses resonant monotop
production (with a scalar resonance of 1 TeV and an invisible fermion of 50~GeV,
all coupling parameters being again set to 0.1).
These event samples have been generated with MG5\_aMC~\cite{%
Alwall:2014hca}, using the NNPDF 2.3 set of parton densities~\cite{Ball:2012cx}
and a UFO model~\cite{Degrande:2011ua} following the monotop model of
Ref.~\cite{Boucheneb:2014wza}. Parton
showering and hadronisation have been simulated with the {\sc Pythia}~8
package~\cite{Sjostrand:2014zea}, and we have included the impact of the
detector response by means of the {\sc Delphes}~3 programme~\cite{%
deFavereau:2013fsa} that internally relies on {\sc FastJet}~\cite{%
Cacciari:2011ma} for object reconstruction and on an appropriate detector
parameterisation describing the performance of the CMS detector during the LHC
Run~1. The validity of our recasting code
has been inferred from a comparison between the \ma~5 and CMS
official results that have been produced from the hard-scattering events that we
have provided to CMS. Event files and generator configuration files can be
obtained from the public analysis database of \ma~5~\cite{pad},
whilst the recasting {\sc C++} code has been additionally submitted to
{\sc InSpire}~\cite{monotop:recast}.

\begin{table}[t]
 \renewcommand\arraystretch{1.2}%
 \begin{center}
  \begin{tabular}{c|c||r|c|c||r|c|c||c}
    & Selection step 
    & CMS & $\epsilon_i^{\rm CMS}$ & $\epsilon_{i, {\rm tot}}^{\rm CMS}$
    & MA5 & $\epsilon_i^{\rm MA5}$ & $\epsilon_{i, {\rm tot}}^{\rm MA5}$
    & $\delta_i^{\rm rel}$\\
    \hline
    0&Nominal                 & 3000  &       &       & 3000 &       & \\
    1&Lepton veto             & 3000  & 1.000 & 1.000 & 2983 & 0.994 & 0.994 &
       0.56\%\\
    2&$p_T(j_1) > 60$~GeV     & 2805  & 0.935 & 0.935 & 2799 & 0.938 & 0.933 &
       0.35\%\\
    3&$p_T(j_2) > 60$~GeV     & 1719  & 0.613 & 0.573 & 1900 & 0.679 & 0.633 &
       10.8\%\\
    4&$p_T(j_3) > 40$~GeV     & 1116  & 0.649 & 0.372 & 1358 & 0.715 & 0.453 &
       10.1\%\\
    5&Veto on the fourth jet  &  598  & 0.536 & 0.200 &  618 & 0.455 & 0.206 &
       15.1\%\\
    6&$M_{3j}<250$~GeV        &  294  & 0.492 & 0.098 &  269 & 0.435 & 0.090 &
       11.5\%\\
    7&$\slashed{E}_T>250$~GeV &   98  & 0.333 & 0.032 &  109 & 0.405 & 0.036 &
       21.6\%\\
    8&$\slashed{E}_T>350$~GeV &   27  & 0.276 & 0.009 &   36 & 0.330 & 0.012 &
       19.9\%\\
    \hline
    S0& 0$b$-jet              &    6  & 0.222 & 0.002 &   12 & 0.333 & 0.004 &
       50.0\%\\
    S1& 1$b$-jet              &   19  & 0.704 & 0.006 &   23 & 0.639 & 0.008 &
       9.2\%
  \end{tabular}
  \caption{\label{tab:fcvalidation}
    Comparison of results obtained with our \ma~5~reimplementation
    (MA5) to those provided by the CMS collaboration (CMS-B2G-12-022) in the
    case of a new physics scenario featuring flavor-changing monotop production.
    The selection and total efficiencies are defined in Eq.~\eqref{eq:epsilons}.
    The official CMS numbers have been derived from the same hard-scattering
    events entering our simulation chain. Those events have been provided to the
    CMS collaboration who accepted to produce an official cutflow.}
 \end{center}
 \renewcommand\arraystretch{1.0}%
\end{table}

\begin{table}[t]
 \renewcommand\arraystretch{1.2}%
 \begin{center}
  \begin{tabular}{c|c||r|c|c||r|c|c||c}
    & Selection step 
    & CMS & $\epsilon_i^{\rm CMS}$ & $\epsilon_{i, {\rm tot}}^{\rm CMS}$
    & MA5 & $\epsilon_i^{\rm MA5}$ & $\epsilon_{i, {\rm tot}}^{\rm MA5}$
    & $\delta_i^{\rm rel}$\\
    \hline
    0&Nominal                 & 4000  &       &       & 4000 &       & \\
    1&Lepton veto             & 4000  & 1.000 & 1.000 & 3989 & 0.997 & 0.997 &
       0.28\%\\
    2&$p_T(j_1) > 60$~GeV     & 3932  & 0.983 & 1.000 & 3947 & 0.989 & 0.986 &
       0.66\%\\
    3&$p_T(j_2) > 60$~GeV     & 2872  & 0.730 & 0.983 & 3044 & 0.771 & 0.761 &
       5.59\%\\
    4&$p_T(j_3) > 40$~GeV     & 1620  & 0.564 & 0.718 & 1944 & 0.639 & 0.486 &
       13.2\%\\
    5&Veto on the fourth jet  &  996  & 0.614 & 0.405 & 1006 & 0.517 & 0.252 &
       15.8\%\\
    6&$M_{3j}<250$~GeV        &  536  & 0.538 & 0.249 &  479 & 0.476 & 0.120 &
       11.5\%\\
    7&$\slashed{E}_T>250$~GeV &  463  & 0.863 & 0.134 &  415 & 0.866 & 0.104 &
       0.30\%\\
    8&$\slashed{E}_T>350$~GeV &  315  & 0.680 & 0.116 &  284 & 0.684 & 0.071 &
       0.59\%\\
    \hline
    S0& 0$b$-jet              &  104  & 0.330 & 0.023 &   90 & 0.317 & 0.023 &
       4.02\%\\
    S1& 1$b$-jet              &  189  & 0.600 & 0.040 &  159 & 0.560 & 0.040 &
       6.69\%
  \end{tabular}
  \caption{\label{tab:resvalidation}
    Same as in Table~\ref{tab:fcvalidation}, but for a new physics scenario
    featuring resonant monotop production.}
 \end{center}
 \renewcommand\arraystretch{1.0}%
\end{table}

The CMS-B2G-12-022 analysis relies on a selection that vetoes the presence of
isolated electrons (muons) with a transverse momentum $p_T$ larger than 10~GeV
(20~GeV) and a pseudorapidity $|\eta|<2.4$ (2.5), where lepton isolation is
imposed by constraining the sum of the transverse momenta of all objects lying
in a cone of radius $R=0.4$ centred on the lepton to be smaller than 20\% of the
lepton $p_T$. The analysis next requires the presence of at most three jets with
transverse momentum $p_T(j_1) >
60$~GeV, $p_T(j_2) > 60$~GeV and $p_T(j_3) > 40$~GeV respectively, and it
additionally forbids the presence of a fourth jet with a $p_T$ greater than
35~GeV. The invariant mass of the three leading jets $M_{3j}$ is then imposed
to be compatible with the top mass $M_{3j}<250$~GeV and the missing energy
has to satisfy $\slashed{E}_T>350$~GeV. Two signal regions S1 and S0 are finally
defined, the difference lying in the presence of either exactly one or exactly
zero $b$-tagged jet.

In Tables~\ref{tab:fcvalidation} and \ref{tab:resvalidation}, we confront
the cutflow charts that have been obtained with \ma~5 to the
official results of CMS for the two benchmark scenarios under consideration.
For each step of the selection, we have calculated the relative ($\epsilon_i$)
and cumulative ($\epsilon_{\rm tot}$) efficiencies, as well as the difference
$\delta_i^{\rm rel}$ between the CMS
official and \ma~5 relative efficiencies, normalised to the CMS
result,
\be
  \epsilon_i = \frac{n_i}{n_{i-1}} \ , \qquad
  \epsilon_{i, {\rm tot}} = \frac{n_i}{n_0}
  \qquad\text{and}\qquad
  \delta_i^{\rm rel} =
     \Big|1 - \frac{\epsilon_i^{\rm MA5}}{\epsilon_i^{\rm CMS}}\Big| \ ,
\label{eq:epsilons}\ee
where $n_i$ and $n_{i-1}$ are the numbers of events after and before the
considered selection step, respectively.
We have found a very good agreement for the resonant monotop benchmark (see
Table~\ref{tab:resvalidation}) reaching a level of 10\%--15\% for all selection
steps. The situation is similar for the flavour-changing scenario case (see
Table~\ref{tab:fcvalidation}), although a 50\% difference between the CMS and the
\ma~5 results is observed for the final S0 selection. This bin is
however populated by a very small number of events so that statistical effects
are likely to be important.

\begin{figure}
  \centering
 \includegraphics[width=.6\textwidth]{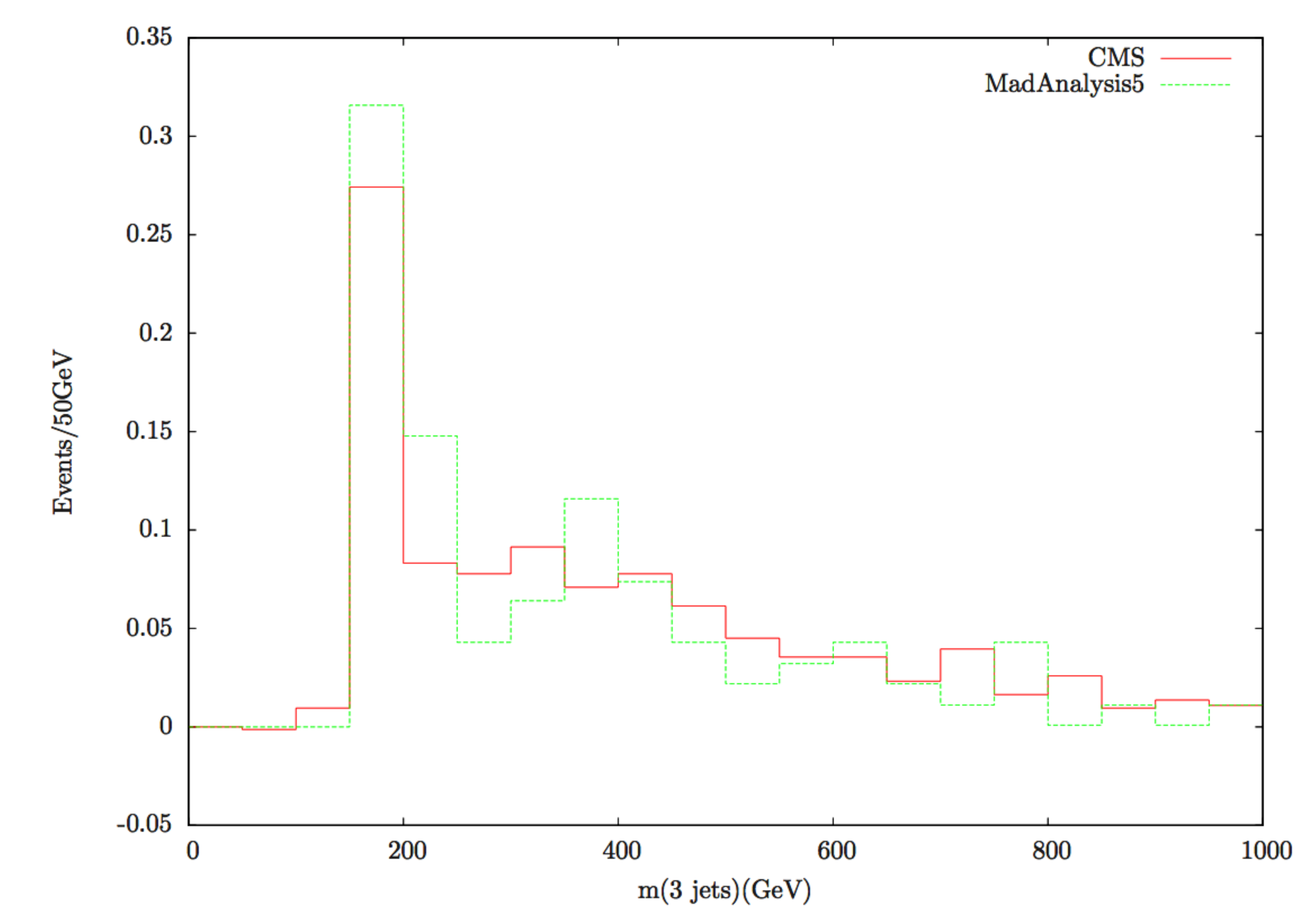}
  \caption{Distribution of the invariant mass of the three leading jets once all
   monotop selection steps have been considered, but the one on $M_{3j}$.
  \label{fig:m3j}}
\end{figure}

We finally also compare the differential distribution in the invariant mass of
the three jets $M_{3j}$ after applying all selections but this one for the
benchmark scenario presented in the CMS analysis note. The latter consists of a
flavour-changing monotop production scenario where an invisible vector state of
mass equal to $700$~GeV is produced in association with a top quark. The results
are shown on Figure~\ref{fig:m3j} and exhibit once again a good agreement.
We therefore consider our reimplementation to be validated.

\begin{figure}
  \centering
  \includegraphics[width=7cm]{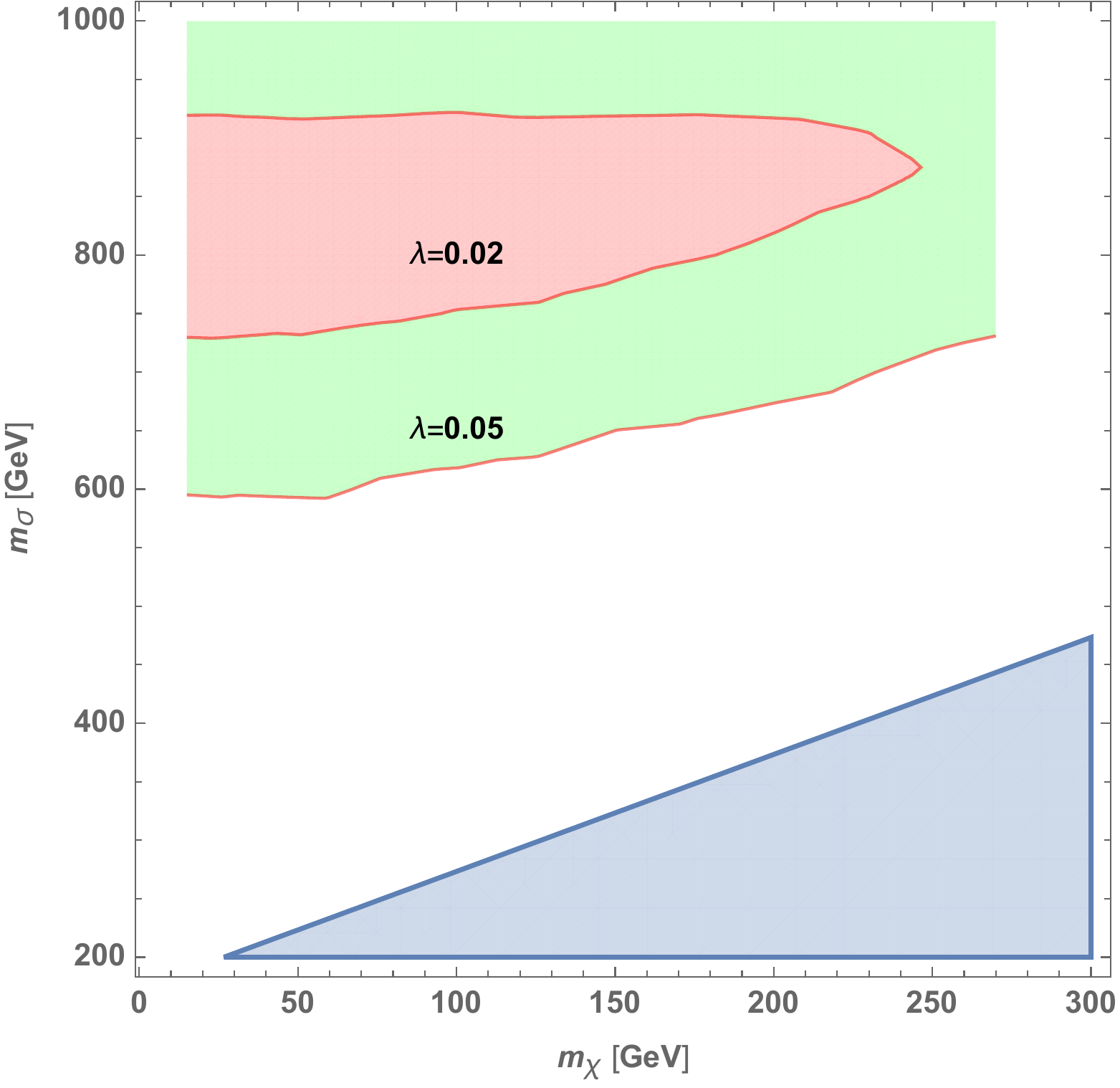}
  \caption{Excluded regions at the 95\% confidence level after reinterpreting
    the results of the CMS-B2G-12-022 analysis in the considered resonant
    monotop model. The results are presented in the $(m_\sigma, m_\chi)$ plane,
    assuming $\mbox{BR}_{t\chi}=100\%$.  The red area corresponds to the
    excluded parameter space region for $\lambda=0.02$, while the green area
    (plus the red one) corresponds to the excluded region for $\lambda=0.05$.
    There is no bound for $\lambda=0.01$ or smaller. Finally, the blue region
    (the triangle area on the bottom right) is not considered as the model
    configuration does not feature any $\sigma$ decay into a monotop system
    $\sigma \to \chi t$. }
\label{fig:monotop}
\end{figure}

In order to extract the constraints on the model parameter space, we generated
the $p p \to \sigma ^{(*)} \to t \chi$ monotop signal for various mass
configurations and extracted, with \ma~5, the number of events $N_s$ populating
the two signal regions S0 and S1. For each signal region, we derived an
exclusion confidence level by generating
$10^5$ Monte Carlo toy experiments in which the actual number of background
events $N_b$ is computed from a Gaussian distribution of mean $\hat N_b$ and
width $\Delta\hat N_b$, the Standard Model background expectation $\hat N_b\pm
\Delta\hat N_b$ being taken from Ref.~\cite{Khachatryan:2014uma}. This allowed
us to calculate the $p$-values
associated with the background-only ($p_B$) and signal-plus-background
($p_{S+B}$) hypotheses. These estimations assume that the number of background
events $N_b$ and signal-plus-background events $N_b+N_s$ are
Poisson-distributed,
and that $N_{\rm data}$ events have been observed (the data numbers being
reported in Ref.~\cite{Khachatryan:2014uma}). We next freely
varied the signal production cross section to the smallest value for which
\be
  1 - \frac{p_{S+B}}{p_B} > 0.95 \ ,
\ee
which corresponds to the cross section $\sigma_{95}$ excluded at the 95\%
confidence level. The final exclusion has been obtained by comparing the most
stringent constraints originating from the two signal regions with
our predicted NLO signal cross section for any given point of the parameter
space. Our results are represented, in the $(m_\sigma, m_\chi)$ plane, in
Figure~\ref{fig:monotop} for two specific $\lambda$ values corresponding to the
typical order of magnitude probed by the other LHC analyses potentially
constraining the model. Heavy mediator masses are usually excluded, provided
that the spectrum is not too compressed. Parameter space configurations in which
the mediator is light cannot however be reached with current data.
Whilst a stronger exclusion could be derived by combining the two regions, we
conservatively ignore this feature as
the combination would require correlation information not shared by CMS.

\subsection{Width constraints}
\label{sec:widths}

As we have seen, the resonantly produced scalar $\sigma$ decays to a monotop
system and to a pair of jets more or less importantly via the relative
magnitude of the couplings $\lambda$ and $y$ respectively. In addition, the $y$
coupling also determines the size of the monotop production cross
section. Maximising both the production rate and the branching ratio into a
monotop system may thus lead to some tensions in the values of the couplings.
The $\sigma$ branching fraction into a $t \chi$ system, in a simplified limit,
only depends on the value of the $\lambda$ and $y$ couplings. Whilst the full
result for the leading order (LO) partial width, $\Gamma(\sigma \to t \chi)$, is
given by
\be
\Gamma(\sigma \to t \chi) = \frac{y^2}{16\pi m_{\sigma}^3} \left( m_{\sigma}^2 -(m_\chi + m_t)^2 \right) \times
\sqrt{m_{\sigma}^4+ (m_\chi^2 - m_t^2)^2-2 m_{\sigma}^2 (m_{\chi}^2 + m_t^2)}\,,
\ee
it simplifies to the simple approximate expression
\be
\Gamma(\sigma \to t \chi)  \simeq \frac{m_\sigma}{16\pi} y^2\,.
\ee
when the $\chi$ and top masses can be neglected. Moreover, in the limit of
massless light quarks, the $\Gamma(\sigma \to \bar{d} \bar{s})$ partial width
reads
\be
  \Gamma(\sigma \to \bar{d} \bar{s}) = \frac{m_\sigma}{2\pi} \lambda^2 \,.
\ee
Combining these two simplified expressions allows to write the $\sigma\to t\chi$
branching fraction as
\be
\mbox{BR}(\sigma \to t \chi)  \simeq \frac{y^2}{y^2+8  \lambda^2} \,,
\ee
where it solely depends on the couplings. These formulas allow to see that, in a
rough approximation, fixing $\lambda$ and requiring a given
$\mbox{BR}(\sigma \to t \chi)$ branching fraction automatically determines the
$y$ coupling. For example if we require $\lambda=0.1$ and $\mbox{BR}(\sigma\to t
\chi)=90\%$ (in order to limit the $\sigma$ decay to dijets, which can give rise
to strong bounds on the model, and keep the number of monotop events that are
expected at the LHC high), a value of $y \simeq 0.85$ is required. This
back-of-the-envelope calculation is not used in the numerical evaluations
performed in this work, where we used exact NLO calculations. Nevertheless it
allows to qualitatively assess the coupling values that are required and their
inter-relations.

\begin{figure}
  \centering
  \includegraphics[width=0.6\textwidth]{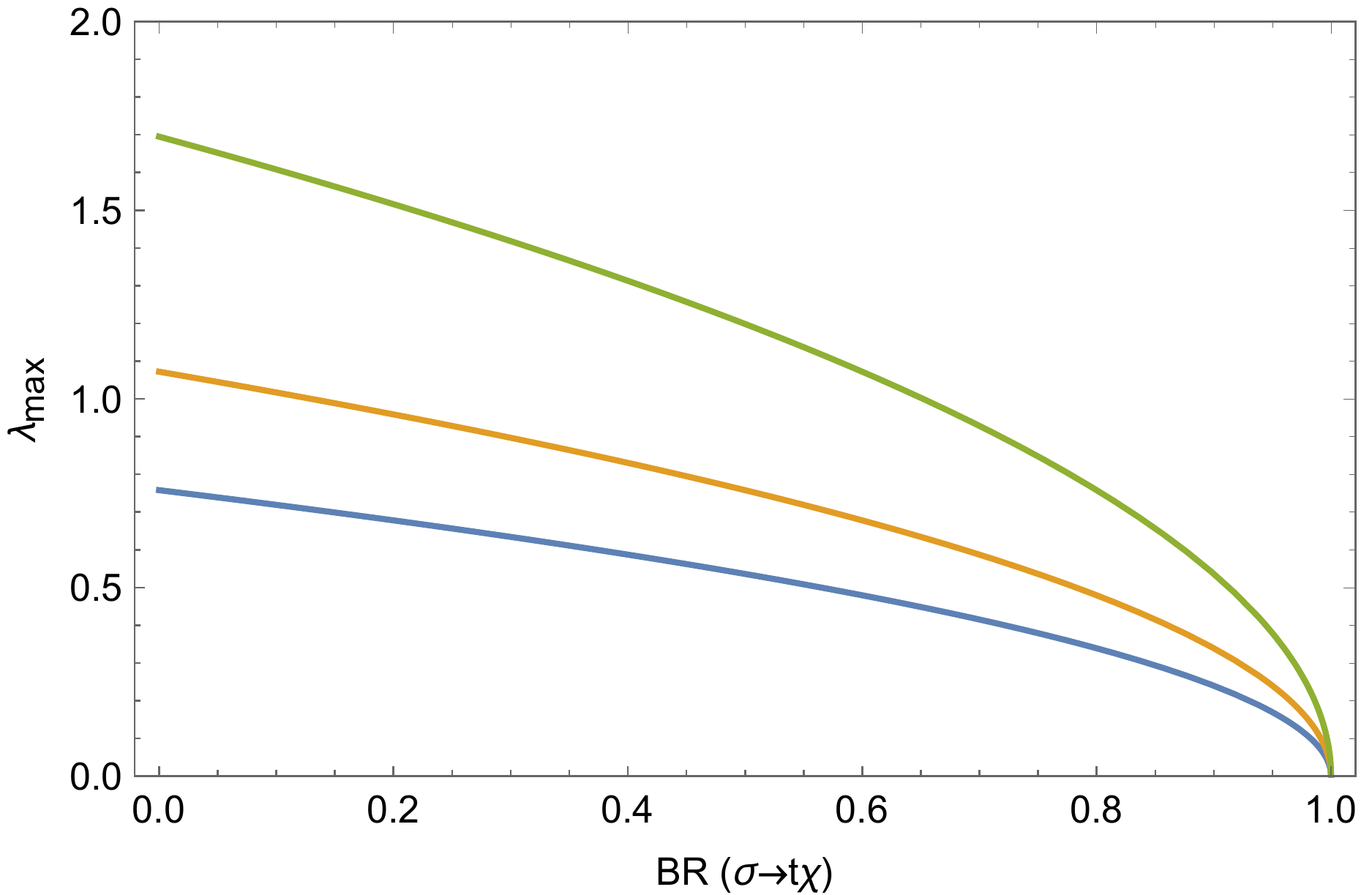}
  \caption{Upper bound on $\lambda$ as a function of $\mbox{BR}_{t\chi}$ to
    obtain $\Gamma_{\rm tot}/m_\sigma$ ratios smaller than 10\% (blue), 20\%
    (orang) and $50\%$ (green).}
\label{fig:NWA}
\end{figure}

The fact that increasing $\lambda$ while keeping the monotop rate fixed forces
us to increase $y$ may lead to parameter space where the total width
of $\sigma$ becomes large. This is however an issue for the reinterpretation of
the experimental searches, as we relied on the narrow width approximation (NWA)
for the simulation of the resonant signal. This therefore imposes an upper bound
on the value of the couplings which may contrast with the requirement of a large
monotop production rate. To quantitatively assess where the NWA breaks down, it is convenient to fix the branching ratio in the monotop channel, and study 
the upper bound on the other coupling $\lambda$ (responsible for the production rate).
The total width of the $\sigma$ resonance, $\Gamma_{\rm tot}$, can thus be
written as
\beq
  \Gamma_{\rm tot} = \frac{\Gamma (\sigma \to \bar{d}\bar{s})}{(1-\mbox{BR}_{t\chi})}\,.
\eeq
The ratio $\Gamma_{\rm tot}/m_\sigma$ depends on the coupling and the
$\mbox{BR}_{t\chi}$ parameter, so that we can 
extract an upper bound on $\lambda$ by imposing the NWA validity.  The limit is mass-independent at LO but a slight mass 
dependence (in $m_\sigma$) is present at NLO. For example varying the  $\sigma$ mass between 320 and 1000 GeV induces
a variation in the allowed maximal value for $\lambda$ smaller than 3\%, while
the difference between the LO and NLO limit generally always lies in the 1--5\%
range.

The results are shown in Figure~\ref{fig:NWA}, where we show contour lines for 
$\Gamma_{\rm tot}/m_\sigma = 10\%$, $20\%$ and $50\%$. For widths larger than $50\%$ of the mass, the $\sigma$ can hardly be thought of as a resonance. 
Also, we see how the larger BR in monotop the smaller $\lambda$ needs to be, thus suppressing the production rates.
For any fixed $\mbox{BR}_{t\chi}$, we can in this way extract an upper bound for
$\lambda$ beyond which the width of the $\sigma$ resonance is too large and
needs to be fully taken into account in the searches.

\begin{figure}
  \centering
  \includegraphics[width=0.6\textwidth]{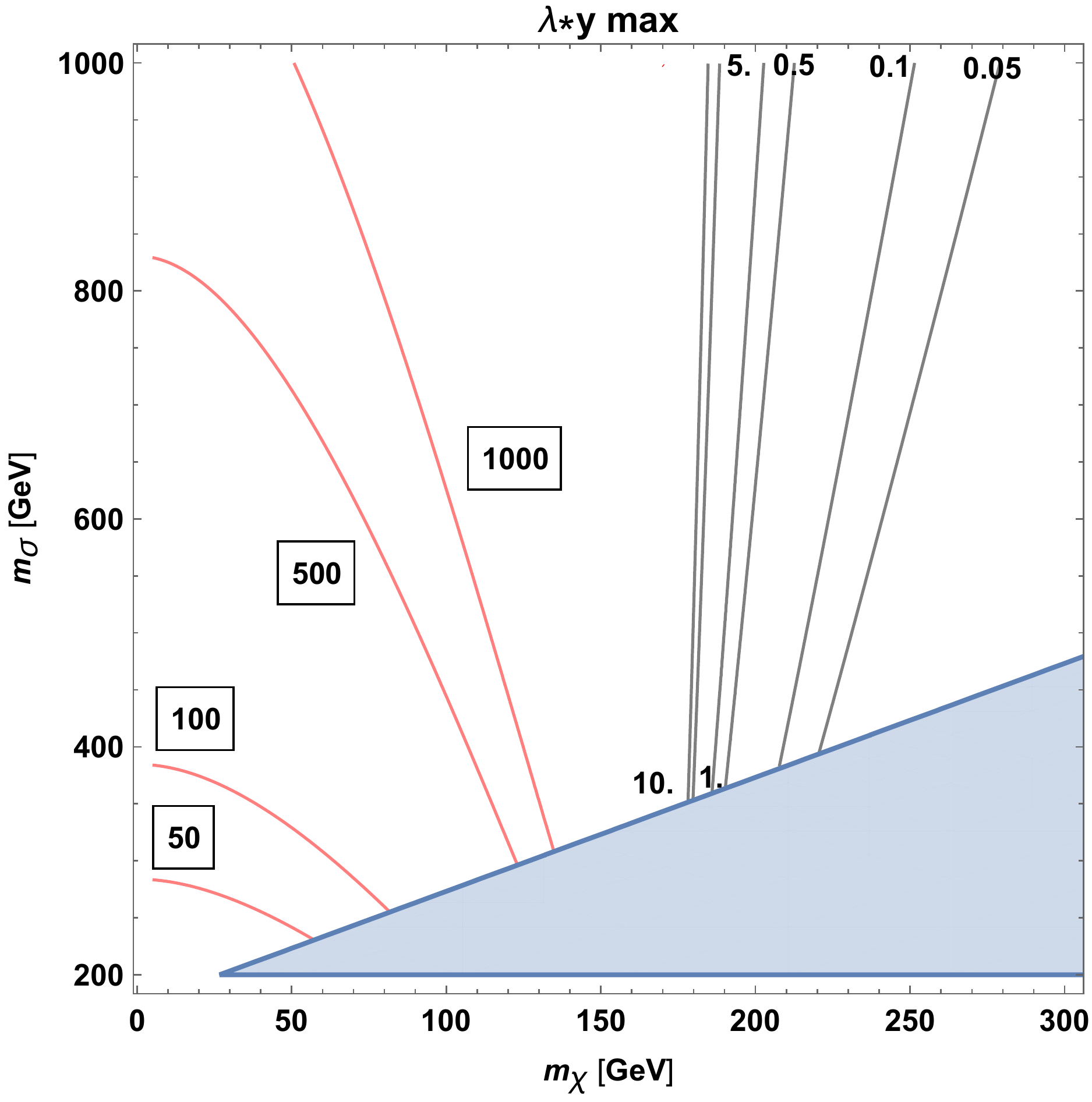}
  \caption{Upper bound on $\lambda \ y$, represented in the $(m_\sigma, m_\chi)$
   plane derived from the width of the $\chi$ fermion and the top quark. The
   black contours correspond to a minimum decay length for the invisible $\chi$
   of 10~metres, while the red ones correspond to a new physics contribution to
   the top width of about 5~GeV.}
\label{fig:widths}
\end{figure}

Another necessary requirement in order to have the monotop signal as the main
feature of the considered new physics model is to make sure that the $\chi$
state produced in association with the top quark is long-lived
enough to escape the detector. Alternatively, one may always assume that $\chi$
dominantly decays into an invisible sector. However, as a fermion, it
can only decay into an invisible fermion plus an invisible boson, or into three
invisible fermions, so that the dark sector has to be rather involved and
one needs to make sure that the couplings to the dark sector are not too large.
Here we focus on the minimal setup in which the only allowed decays of $\chi$
derive from the two couplings included in the Lagrangian of Eq.~\eqref{eq:mod}.
We distinguish three kinematic regimes:
\begin{enumerate}
\item for $m_\chi > m_{\rm top}$, the three-body decay $\chi \to t d s$ is open,
  thus potentially giving very strong bounds on the couplings;
\item for $m_W < m_\chi < m_{\rm top}$, the decay is four-body, $\chi \to W^+ b
  d s$ and takes place via a virtual top quark;
\item for $m_b < m_\chi < m_W$, only a five-body decay is allowed via both an
  off-shell top quark and $W$-boson.
\end{enumerate}
The issue of the width of the $\chi$ fermion has been studied in Ref.~\cite{%
Wang:2011uxa}, where it has been showed that for masses of a few GeV below the
$W$-boson mass $m_W$, decay lengths in the metre range can be obtained.
For all three kinematic regimes defined above, the decay proceeds through a
virtual $\sigma$ exchange, and the width is proportional to the product of the
two couplings $(\lambda \ y)^2$. By requiring that the decay length of $\chi$ is
larger than the typical scale of an LHC detector, {\it i.e.} 10 metres, we can
obtain an upper  bound on $\lambda \ y$, as shown in Figure~\ref{fig:widths}.

Moreover, for $m_\chi < m_{\rm top}$, the same $\sigma$-exchange induces a
three-body decay for the top quark, $t \to \chi \bar{d} \bar{s}$. The
corresponding partial width, which is also proportional to $(\lambda \ y)^2$,
can be constrained by the direct measurement of the top width by the CDF
collaboration at the TeVatron~\cite{Aaltonen:2013kna},
\be
  \Gamma_t < 6.38~\text{GeV at the 95\% confidence level}.
\ee
Assuming that the Standard Model width is unaffected ($\Gamma_t^{\rm SM} =
1.41$~GeV), this leads to a bound on the $\Gamma (t \to  \chi \bar{d} \bar{s})$
partial width,
\be
  \Gamma (t \to  \chi \bar{d} \bar{s}) \lesssim 5~{\rm GeV}.
\ee
Whilst more precise determinations of the top width exist, as for instance in
Ref.~\cite{Khachatryan:2014nda}, these indirect measurements all assume the
absence of non-standard decay channels and cannot thus be used here.
The above bound is represented by red contours in Figure~\ref{fig:widths}.
Notably, it is complementary to the $\chi$ width bounds, as it is more sensitive
to the low $m_\chi$ region. Nevertheless it does not impose very strong limits
on the couplings.

\section{Discussion and conclusions}
\label{sec:LHC}
\begin{figure}
  \centering
  \includegraphics[width=0.6\textwidth]{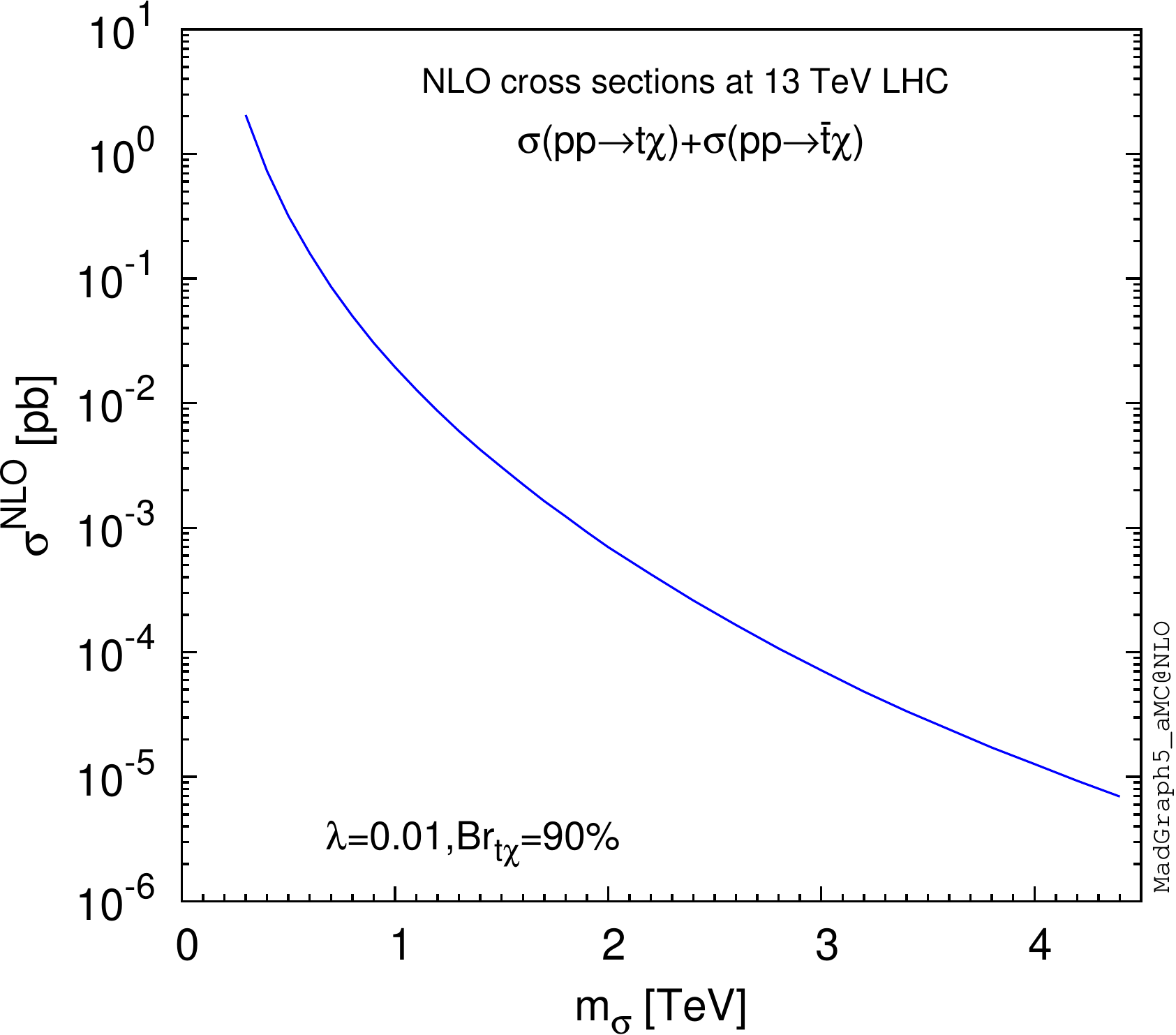}
  \caption{NLO monotop production cross sections at the LHC operating at a
    centre-of-mass energy $\sqrt{s}=13$~TeV for different $m_{\sigma}$ values,
    assuming $m_{\chi}=50$~GeV, $\lambda=0.01$, ${\rm BR}_{t\chi}=90\%$ and
    $y \approx 0.0977$. Varying the mass of $m_{\chi}$ from a few GeV to 100~GeV
    barely affects the value of ${\rm BR}_{t\chi}$, which is a multiplicative
    factor entering the monotop production cross section.}
\label{fig:monotop13TeV}
\end{figure}

At 13 TeV one can have an idea of the potential of the LHC to observe a
monotop signal by considering the production cross section in parameter space
regions still allowed by data. A detailed study is not possible at present as
this would require the generation of the corresponding Standard Model
background, and even this would be only a rough analysis, considered that the
13~TeV environment and background can only be accurately determined using real
data. Nevertheless, the signal cross section ranges from a few picobarns for
$m_{\sigma}=300$ GeV to the femtobarn level for resonance masses lying around
$m_{\sigma}=2$ TeV, as illustrated in Figure~\ref{fig:monotop13TeV} for a
specific set of new physics couplings and masses. Those large
numbers could in principle motivate the experimental collaborations to attempt a
monotop search aiming to discover (or bound in a less optimistic case) the
corresponding signal at the LHC Run 2.

We have provided in this work elements allowing to assess this question
stronger. We have collected bounds of different origins that range from specific
searches at colliders that we have recasted to reinterpret their results in the
context of the model under consideration, to limitations implied by the presence
of the monotop signal as a key new physics model feature. In the rest of this
section, all these bounds will be put together in order to determine which parts
of the parameter space are still open and if a monotop signal visible at Run~2
could be expected. On top of the ``vanilla" monotop model in which the
${\rm BR}_{t\chi}$ parameter is close to 1, we have also considered deviations
in which monotop production should be substantial, with the value of
${\rm BR}_{t\chi}$ being lowered to 90\%, 75\%, 50\% and 10\%. However, in this
more general case, a global and detailed analysis including dijet resonance
searches, new physics searches in the top quark with missing energy and jets
channel as well as in the multijet mode need to be considered more deeply.

\begin{figure}
  \centering
  \includegraphics[width=0.48\columnwidth]{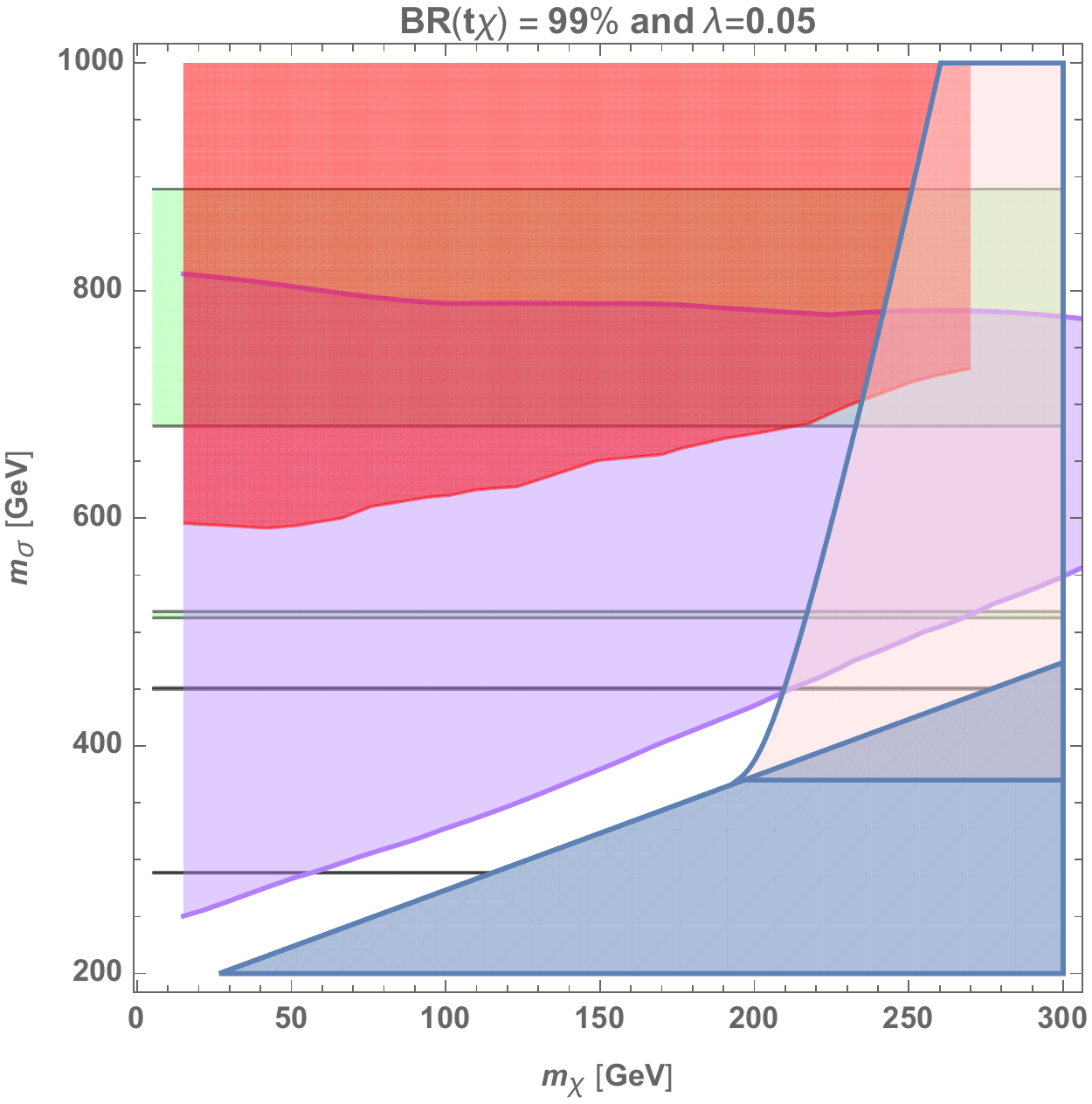}
  \includegraphics[width=0.48\columnwidth]{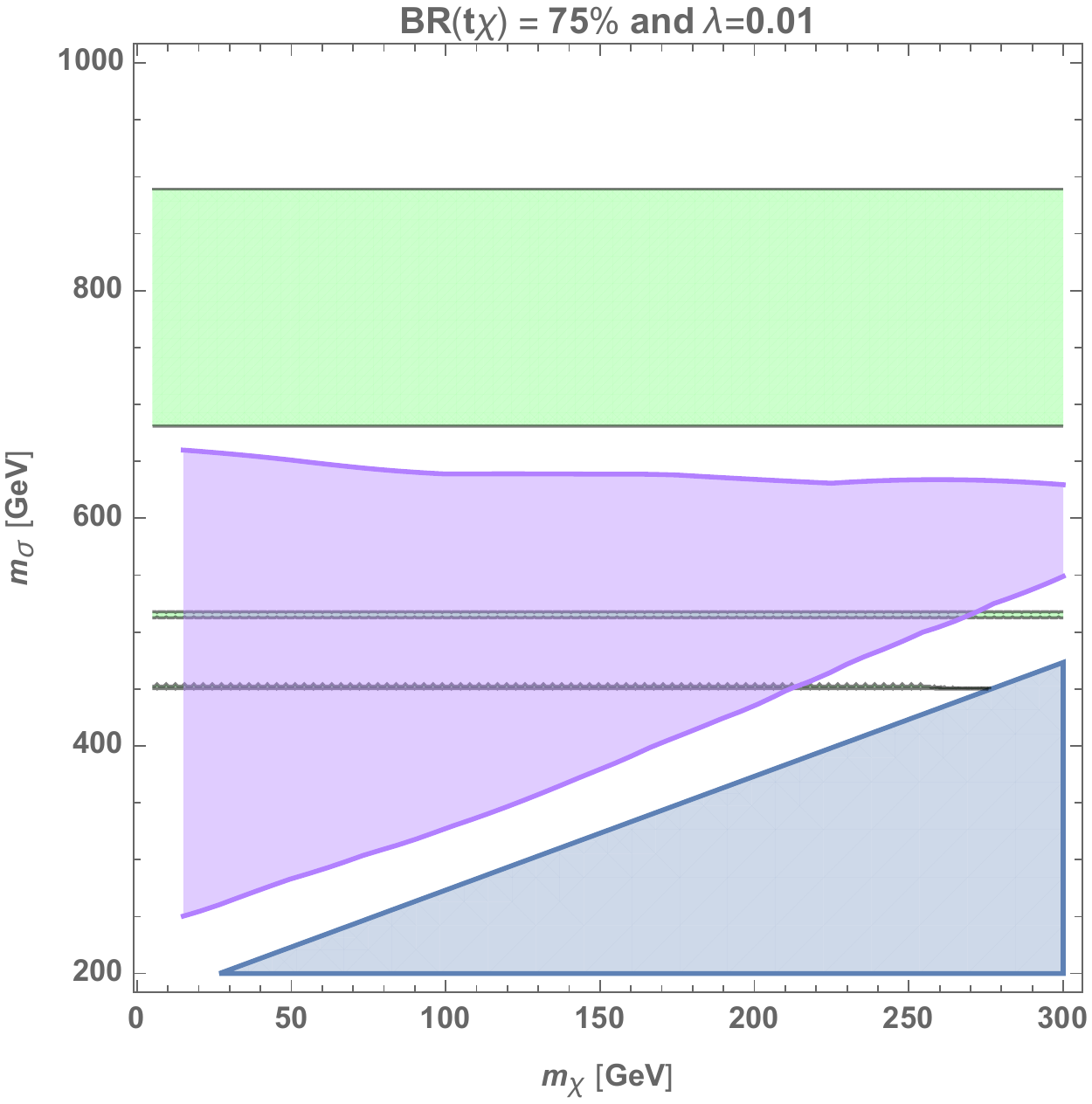}\\
  \includegraphics[width=0.48\columnwidth]{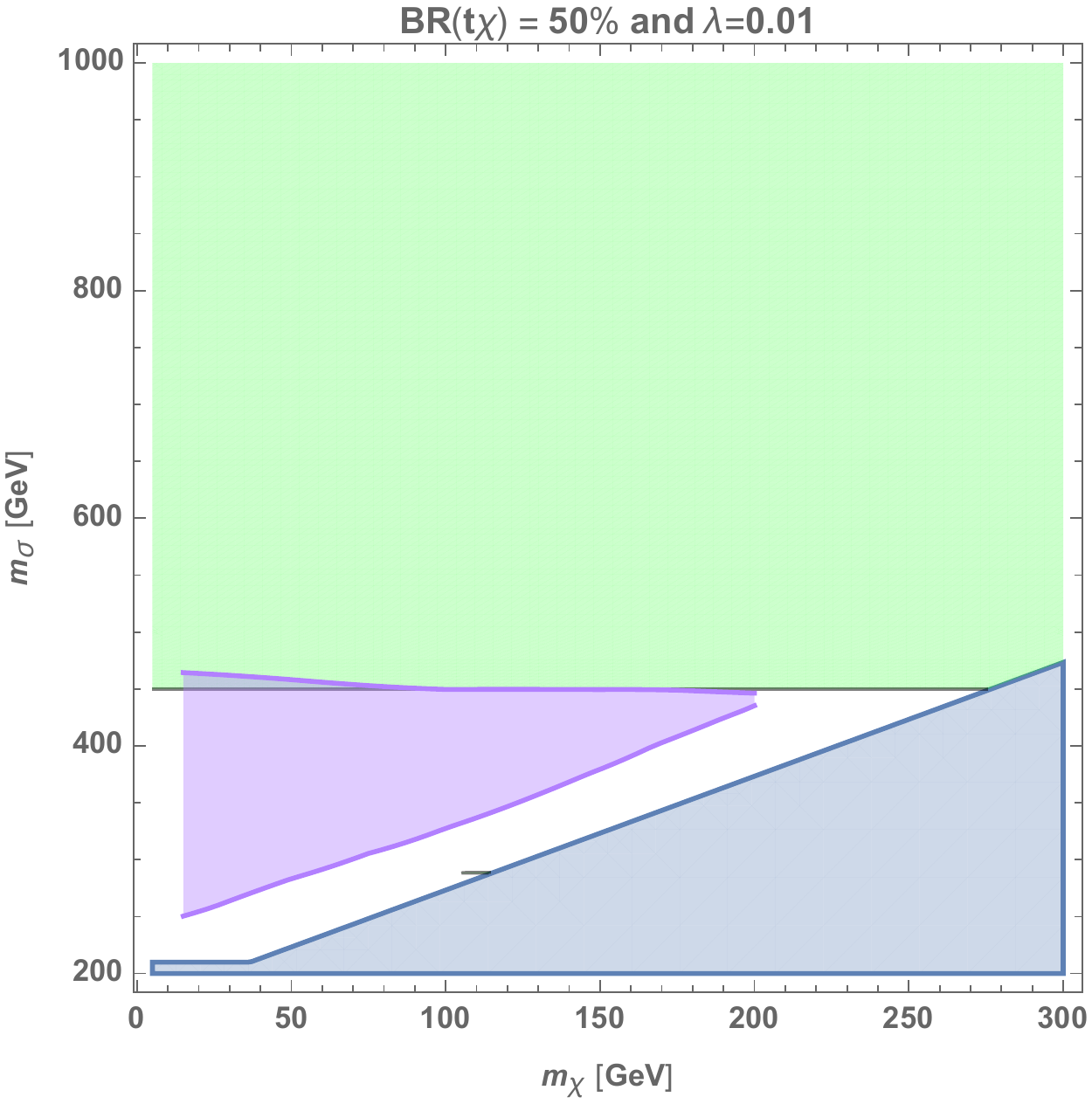}
  \includegraphics[width=0.48\columnwidth]{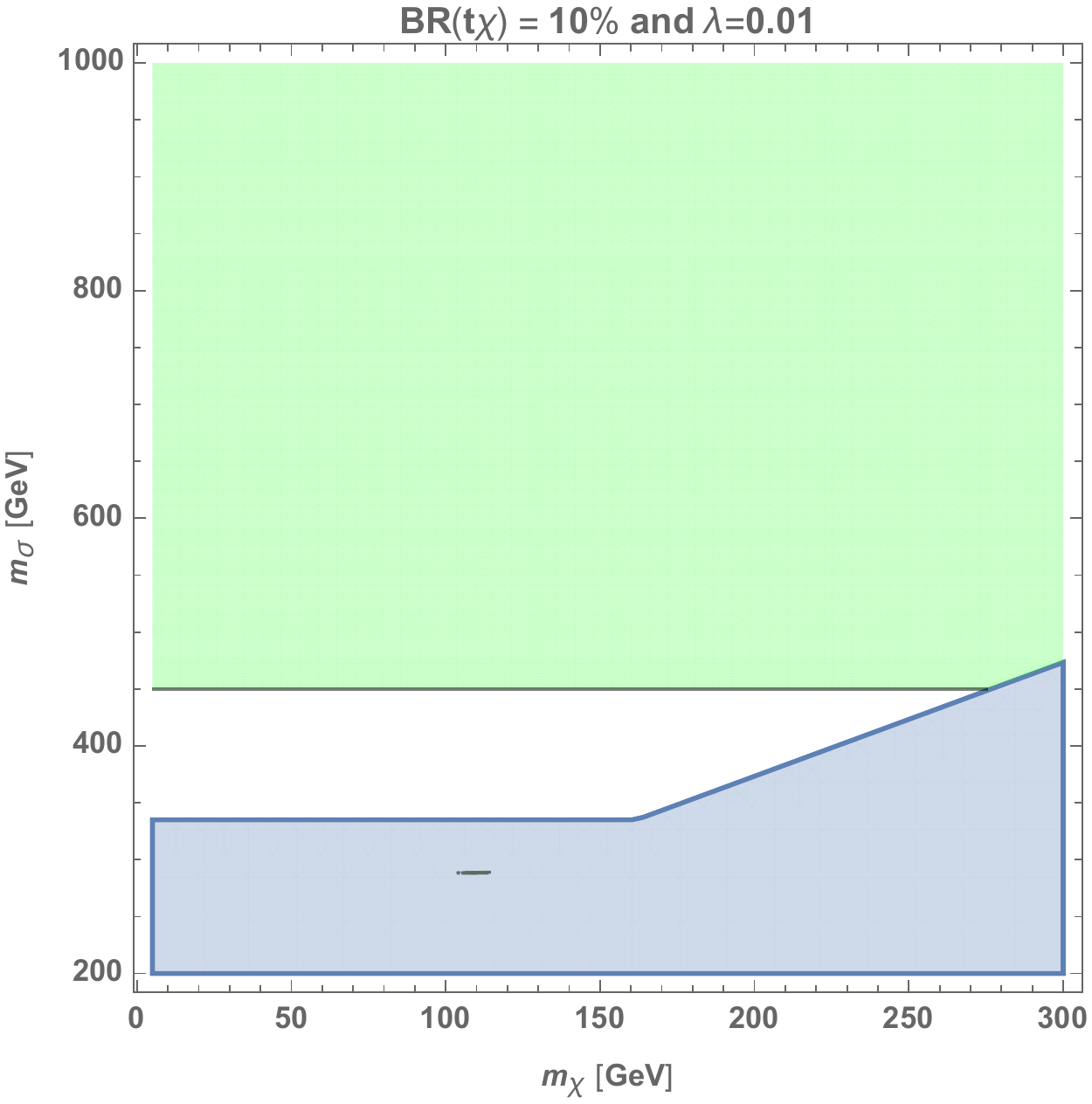}
  \caption{Current constraints on the parameter space of the considered model
    featuring resonant monotop production at colliders. The results are
    presented in $(m_\chi, m_\sigma)$ planes for various $\lambda$ and
    ${\rm BR}_{t\chi}$ values. Whilst the white region are allowed, the coloured
    areas are excluded by at least one of the bounds described in
    Section~\ref{sec:constraints}. We refer to the text for more details.}
  \label{fig:BR}
\end{figure}

In Figure~\ref{fig:BR}, we present the bounds on the parameter space in the
$(m_{\chi}, m_{\sigma})$ plane for four choices of the $\sigma$ decay rates into
a monotop system, ${\rm BR}_{t\chi}=$ 99\% (upper left panel), 75\% (upper right
panel), 50\% (lower left panel) and 10\% (lower right panel). The results
include constraints originating from all the searches and features discussed in
Section~\ref{sec:constraints}, namely stop searches, dijet searches, monotop
searches and the decay length of the $\chi$ fermion that must be larger than
10~metres. Moreover, we restrict the chosen values of the $\lambda$ parameter so
that the NWA for $\sigma$ is valid. Imposing $\Gamma_{tot}/m_\sigma< $ 20\%, we
obtain from Figure~\ref{fig:NWA},
\be
  \lambda < 0.35,\ 0.56,\ 0.79\  \text{and}\ 1.06\qquad\text{for}\qquad
  \mbox{BR}_{t\chi}=99\%,\ 75\%,\ 50\%\  \text{and}\ 10\%\ \text{respectively}.
\ee
The figures
include both $\lambda$-independent and $\lambda$-dependent bounds, so that we
indicate the chosen $\lambda$ value in the upper caption of each subfigure. We
moreover recall that, for each of the considered constraint, theoretical
predictions have been achieved at the NLO accuracy in QCD thanks to recent
developments at the level of the Monte Carlo simulations. This could serve as
a starting ground for more detailed analyses to be performed by the LHC
experiments at Run 2.

The blue regions (triangle areas on the bottom right of each subfigure)
correspond to parameter space configurations in which a monotop signal cannot be
produced resonantly, the $\sigma \to \chi t$ decay being kinematically closed. We therefore
omit it from our analysis. We also represent by rectangular blue areas (bottom
left of the two lower subfigures) the regions that are excluded by resonance
search in the dijet-pair channel and that we have studied in
Figure~\ref{fig:dijetpair}. As already found in Section~\ref{sec:dijetpairs},
the most powerful searches concern CMS and ATLAS analyses of 13 TeV LHC data,
and they only have some constraining power for low $\sigma$ masses and a large
branching ratio associated with the $\sigma\to jj$ decay ({\it i.e.} a not too
large $\mbox{BR}_{t\chi}$ value). The light violet regions are excluded by stop
searches, as detailed in Section~\ref{sec:stops}. Typically, the limits
presented in Figure~\ref{fig:stoppair} are rescaled down proportionally to the
decreasing value of $\mbox{BR}_{t\chi}$ (that lowers the signal production
rate). All figures then feature black horizontal lines that delimitate the green bands
of the mass planes that are excluded by dijet searches (see
Section~\ref{sec:dijet}). For $m_\sigma$ in the 300--450~GeV mass window, dijet
bounds are weak as this corresponds to a mass configuration only probed by
TeVatron searches. In contrast, LHC searches are sensitive for other $\sigma$
masses and are especially stronger for masses lying in the 200--300~GeV and
450--1000~GeV ranges. Whilst the white areas are in principle reachable by dijet
searches, the chosen $\lambda$ values are too small to yield any constraint. In
the upper left panel of the figure (for which ${\rm BR}_{t\chi}= 99\%$), the
monotop bounds presented in Figure~\ref{fig:monotop} are overlaid, so that the
upper part of the mass plane is excluded (dark red region). Those searches
however quickly lose sensitivity with smaller $\mbox{BR}_{t\chi}$ values in
association with a $\lambda$ value also five times smaller.
Finally, this panel also exhibits a trapezoid pink area that corresponds to a
region in which the decay length of the $\chi$ fermion is smaller than 10
metres, so that there is actually no monotop signal in there. There is no
bound for any of the the three other $(\lambda, \mbox{BR}_{t\chi})$
configurations, as the smaller $\lambda$ value ensures a larger decay length.

To summarise, current limits severely restrict the model parameter space for
light new physics states. Only small specific subregions are still allowed by
data, and it is clear that future results will allow one to draw conclusive
statements. In contrast, the heavier cases are still viable.

\acknowledgments
We are grateful to Josselin Proudom and Paolo Torrielli for useful discussions
and comments in the early stage of this work, to Nishita Desai, Stefan Prestel
and Peter Skands for comments on the color junction implemented in
{\sc Pythia}~8 and to Ivan Mikulec and Cristina Su\'arez for clarifying details
on the CMS exclusion limits of Ref.~\cite{Sirunyan:2017nvi}. We also thank Jun
Guo and Khristian Kotov for their help in recasting the Run~1 CMS monotop
analysis. AD and GC acknowledge partial support from the Labex-LIO (Lyon
Institute of Origins) under grant ANR-10-LABX-66, FRAMA (FR3127, F\'ed\'eration
de Recherche ``Andr\'e Marie Amp\`ere''). BF and HSS are supported by French
state funds managed by the Agence Nationale de la Recherche (ANR), in the
context of the LABEX ILP (ANR-11-IDEX-0004-02, ANR-10-LABX-63).

\bibliographystyle{JHEP}
\bibliography{bibliography}

\end{document}